\documentclass[nofootinbib,aps,prd,12pt]{revtex4-1}
\pdfoutput=1
\usepackage{graphicx}
\usepackage{cancel}
\usepackage{amssymb}
\usepackage{amsmath}
\usepackage{amsfonts}
\usepackage{bm}
\usepackage{times}
\usepackage{color}
\usepackage{mathptmx}
\usepackage[normalem]{ulem}
\usepackage{ mathrsfs }
\usepackage{textcomp}
\usepackage{slashed}
\usepackage{booktabs}
\usepackage{latexsym}
\usepackage{verbatim}
\usepackage{extarrows}
\usepackage{multirow}
\usepackage{xcolor}
\usepackage{epsfig}
\usepackage{epstopdf}
\usepackage{autobreak}
\usepackage{subfigure,psfrag}
\usepackage[pdfpagelabels]{hyperref}
\hypersetup{
	colorlinks,
	linkcolor={black!50!black},
	citecolor={blue!50!black},
	urlcolor={blue!80!black},
	bookmarksopen=true
}
\usepackage{anyfontsize}
\graphicspath{{./Figures/}}
\newcommand{\hp}{{\it H5plane}}
\newcommand{\lm}{{\it low mass benchmark}}

\makeatletter
\renewcommand*\env@matrix[1][\arraystretch]{%
	\edef\arraystretch{#1}%
	\hskip -\arraycolsep
	\let\@ifnextchar\new@ifnextchar
	\array{*\c@MaxMatrixCols c}}
\makeatother
\def\figureautorefname~#1\null{Fig.\,#1\null}
\def\tableautorefname~#1\null{Tab.\,#1\null}

\def\equationautorefname~#1\null{Eq.\,(#1)\null}

\begin{document}

\title{Gravitational wave and Collider searches for the EWSB patterns }


\author{Ligong Bian $^{1,2}$}
\email{lgbycl@cqu.edu.cn}

\author{Huai-Ke Guo$^{3}$}
\email{ghk@ou.edu}

\author{Yongcheng Wu$^{4}$}
\email{ycwu@physics.carleton.ca}

\author{Ruiyu Zhou $^{1}$}
\email{zhoury@cqu.edu.cn}

\affiliation{
	$^1$~Department of Physics, Chongqing University, Chongqing 401331, China
	\\
    $^2$ Department of Physics, Chung-Ang University, Seoul 06974, Korea\\
    $^3$ Department of Physics and Astronomy, University of Oklahoma, Norman, OK 73019, USA\\
    $^4$ Ottawa-Carleton Institute for Physics, Carleton University, 1125 Colonel By Drive, Ottawa,
Ontario K1S 5B6, Canada
}
\date{\today}

\begin{abstract}
We study the Electroweak symmetry breaking mechanism with extra Electroweak symmetry breaking contributions (eEWSB) that are bounded by the 
Fermi constant and limits from the related collider searches.
The eEWSB is helpful to build a different zero temperature vacuum structure from the Standard Model (SM), and therefore leads to different Electroweak phase transition patterns at the early Universe. We investigate the collider search prospects and gravitational waves (GW) predictions from the strongly firstly order phase transition (SFOEWPT) in this scenario. The Higgs pair searches at lepton colliders are found to be complementary with the GW searches of the SFOEWPT parameter spaces.
\end{abstract}

\pagenumbering{Alph}
\begin{titlepage}
\maketitle
\thispagestyle{empty}
\end{titlepage}
\baselineskip=16pt
\pagenumbering{arabic}

\vspace{1.0cm}
\tableofcontents
\thispagestyle{empty}
\newpage

\section{Introduction}

The observation of the SM Higgs at 126 GeV at LHC~\cite{Aad:2012tfa,Chatrchyan:2012xdj} is a milestone of the particle physics, 
which means that the W and Z boson obtain their masses through the Electroweak symmetry breaking (EWSB) mechanism. 
The cubic and quartic Higgs couplings are supposed to be crucial to reveal the Higgs potential shape and the EWSB mechanism. The sensitivity of measurement of 
these couplings at LHC is pretty low, while future precision measurements are able to tell if there are new physics beyond the Standard Model (SM) that could drive deviation of the EWSB and how large the deviation could be.
The observation of gravitational waves from the Binary Black hole merger by the LIGO and Virgo collaborations~\cite{Abbott:2016blz} opens a new era to search for 
fundamental physics. An important category of gravitational waves is a stochastic background~\cite{TheLIGOScientific:2016dpb} 
originated from the earth Universe.
One important source of this kind is a strongly first order Electroweak phase transition (SFOEWPT), which gives a dynamical explanation of the EWSB as the Universe cools down, and is a crucial ingredient in the explanation of the baryon asymmetry of the Universe within the Electroweak baryogenesis mechanism.\footnote{The SFOEWPT is one of the three Shakharov conditions~\cite{Sakharov:1967dj} that quenches the sphaleron process inside the bubble and therefore preserve 
the baryon asymmetry  being generated (see Ref.~\cite{Morrissey:2012db} for a recent review on Electroweak baryogenesis).} 

New physics that takes part in the Electroweak phase transition process may or may not contribute an extra component of EWSB contribution. For example, the SM plus real singlet model (xSM) has been extensively studied where the singlet scalar do not contribute any EWSB contribution(see Ref.~\cite{Alves:2018oct} for a recent study). Meanwhile, the triplets in the Georgi-Machacek (GM) model, as will be studied in this work, can indeed contribute to the EWSB. Though both of the two models share the same vacuum structure topology, the triplets contribution to the EWSB, i.e. the extra EWSB contribution, is bounded by the Fermi constant and gauge boson related collider searches.
Therefore, one can expect different collider phenomenology, different SFOEWPT behavior and thus different gravitational wave signal predictions for different amount of the extra EWSB contributions. 

The zero temperature vacuum structure with extra local minimum in addition to the Electroweak vacuum could yield the possibility of multi-step phase transition as well as one-step phase transition.  
In Ref.~\cite{Dorsch:2017nza,Harman:2015gif}, the relation between the zero temperature potential difference and the SFOEWPT condition has been studied within the 2HDM. For previous studies of multi-step phase transition and related vacuum structure at zero temperature, we refer to Ref.~\cite{Jiang:2015cwa,Bian:2017wfv,Bernon:2017jgv,Cheng:2018axr,Bian:2018mkl,Bian:2018bxr,Chao:2017vrq,Cheng:2018ajh,Alves:2018oct}. In this work, 
we use the Georgi-Machacek model to reveal that, the extra EWSB contribution can 
induce one-step or two-step SFOEWPT depending on the vacuum structure that has been studied previously by us in Ref~\cite{Zhou:2018zli}. The one-step SFOEWPT occurs with the symmetry change from $SU(2)_L\times SU(2)_R$ to the phase where Electroweak symmetry is broken.
The two-step SFOEWPT occurs with the first-step being the symmetry change of $SU(2)_L\times SU(2)_R\to SU(2)_V$, and the following second step being the dynamical broken of the Electroweak symmetry.   
In this work we improved the algorithm for the calculation of the critical order parameters of the phase transition. 
We further evaluate the gravitational wave signals being generated during the SFOEWPT. 
In comparison with the one-step situation, the gravitational wave signal spectrum generated from two-step SFOEWPT is found much easier to be probed by the projected space-based interferometers, such as: LISA~\cite{Klein:2015hvg},
BBO, DECIGO (Ultimate-DECIGO)~\cite{Kudoh:2005as}, TianQin~\cite{Luo:2015ght} and Taiji ~\cite{Gong:2014mca} programs. This is significantly different from the xSM case as was studied in Ref.~\cite{Alves:2018oct}, where the vacuum expectation value (VEV) of the extra singlet is more free from the limits of the collider searches.

This work is organized as follows: The vacuum structure analysis and the phase transition calculation approach are given in~\autoref{sec:EWPTdy}. In~\autoref{sec:SFOEWPT}, we show the relation between the phase transition and the collider phenomenology, and demonstrate how these two interplay on the extra EWSB contributions. The gravitational wave signal predictions from the one-step and two-step SFOEWPT are investigated in~\autoref{sec:GWTn}. The collider search prospects for the SFOEWPT valid regions are addressed in~\autoref{sec:cols}. We finally conclude with~\autoref{sec:conc}. Some details about the model are listed in Appendix.

\section{The EWPT dynamics and methodology }
\label{sec:EWPTdy}

In this section, we first develop the methods for vacuum structure analysis and phase transition critical order parameter analysis, which can be applied to phase transition analysis of any multi-scalar models, such as xSM, 2HDM, 2HDM$+$S, NMSSM, etc. In the GM model, at the critical temperature, the strongly first order phase transition condition could be fulfilled when $v_c/T_c\equiv \sqrt{h_\phi(T_C)^2+8h_\xi(T_C)^2}/T_C\geq 1$ \cite{Chiang:2014hia,Zhou:2018zli}.  

\subsection{On the vacuum structures and the possible EWPT patterns}

\begin{figure}[!htp]
\begin{center}
\includegraphics[width=0.5\textwidth]{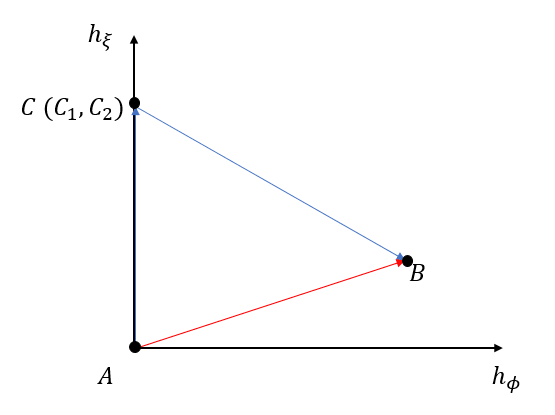}
\end{center}
\caption{The vacuum structure in GM model.}
\label{fig:VSGM}
\end{figure}

The leading order zero temperature effective potential of the GM model is~\cite{Zhou:2018zli} 
\begin{eqnarray}
V_{0}&=& \frac{1}{4}(4 h_{\phi}^{4} \lambda_1 + 2 (h_{\xi}^2+h_{\chi}^2)(m_2^2+2\lambda_2 (h_{\xi}^2+h_{\chi}^2))+2\lambda_3(2 h_{\xi}^4 + h_{\chi}^4)\nonumber\\
&+& h_{\phi}^2 (2 m_1^2 + 4  \lambda_4 h_{\xi}^2 + h_{\xi}(2\sqrt{2} \lambda_5 h_{\chi}  + \mu_1) + h_{\chi}(4 \lambda_4   h_{\chi}+\lambda_5  h_{\chi} + \sqrt{2} \mu_1)) + 12 \mu_2  h_{\xi} h_{\chi}^2 )\;.
\label{eq:Vtree}
\end{eqnarray}
For the vacuum structure studies, we impose $h_\chi=\sqrt{2}h_\xi$ as required by the custodial symmetry, which ensures $\rho=1$ at leading order. The general vacuum structure determined by the above potential is shown in~\autoref{fig:VSGM}, where A is the $(h_\phi,h_\xi)=(0,0)$ vacuum, B is the desired EW vacuum, $C_{1,2}$ are the alternative vacuums with $h_\phi=0$ (the $SU(2)_V$ vacuum). In this paper we consider the case where there are two possible C points in GM model which can be expressed as below,
\begin{align}
C_{1(2)}~point~:~h_{\phi} \to 0~,~h_{\xi} \to \frac{-3 \mu_2 \pm \sqrt{-12m_2^2 \lambda_2-4m_2^2 \lambda_3+9\mu_2^2}}{4(3\lambda_2+\lambda_3)}\;.
\label{equ:eqvac}
\end{align}

The scalar potential at the EW vacuum (B) should be the global minimum one and the value of the scalar potential at the original point 
is the maximal one of these three. The scalar potential at these three different vacuum points, $V_0(A),~V_0(B),~V_0(C_{1(2)})$, are
\begin{align}
V_0(A) =&~0\;, \nonumber\\
V_0(B) =& - \lambda_1 \nu_\phi^4 - 3 \nu_\xi^3 (\mu_2 + (3\lambda_2+\lambda_3)\nu_\xi) - \frac{3}{8} \nu_\xi(\mu_1+4(2 \lambda4+\lambda5)\nu_\xi)\nu_\phi^2\;,\nonumber\\
V_0(C_{1(2)}) =& -\frac{3}{256 \nu_{\xi}  (3 \lambda_2+\lambda_3)^3}( \textit{F} \mp 3 \mu_2)^2 (\mu_2 (24 \nu_{\xi} ^2 (3 \lambda_2+\lambda_3) \mp 2 \nu_{\xi}\textit{F})\nonumber\\
&+(3 \lambda_2+\lambda_3) (16 \nu_{\xi} ^3 (3 \lambda_2+\lambda_3)+\nu_{\phi}^2 (4 \nu_{\xi}  (2 \lambda_4+\lambda_5)+\mu_1))\; \nonumber\\
&+6 \mu_2^2 \nu_{\xi} )\;, \\
\rm{where}\nonumber\\
\textit{\it{F}}=&\bigg(\frac{\nu_{\phi} ^2 (3 \lambda_2+\lambda_3) (4 \nu_{\xi}  (2 \lambda_4+\lambda_5)+\mu_1)}{\nu_{\xi} }+(4 \nu_{\xi}  (3 \lambda_2+\lambda_3)+3 \mu_2)^2\bigg)^{1/2}\;.
\end{align}

The one-step phase transition would take place when $V_0(A)>V_0(B)$ ($\Delta V_0(AB)>0$) and $-12m_2^2 \lambda_2-4m_2^2 \lambda_3+9\mu_2^2 < 0$.
Meanwhile, the two-step phase transition might happen when $-12m_2^2 \lambda_2-4m_2^2 \lambda_3+9\mu_2^2 \geq 0$ and $V_0(A)>V_0(C_{1(2)})>V_0(B)$ (with $\Delta V_0(AC_{1(2)})>0,\Delta V_0(C_{1(2)}B)>0$). The potential differences are given as
\begin{eqnarray}
\Delta V_0({AB})&\equiv& V_0(A)-V_0(B) \nonumber\\
&=&\lambda_1 \nu_\phi^4 + 3 \nu_\xi^3 (\mu_2 + (3\lambda_2+\lambda_3)\nu_\xi) + \frac{3}{8} \nu_\xi(\mu_1+4(2 \lambda_4 \nonumber\\
&+&\lambda_5)\nu_\xi)\nu_\phi^2\;,\\
\Delta V_0({AC_{1(2)}})&\equiv&V_0(A)-V_0(C_{1(2)}) \nonumber\\
&=& \frac{3}{256 \nu_\xi  (3 \lambda_2+\lambda_3)^3}( \textit{F} \mp 3 \mu_2)^2 (\mu_2 (24 \nu_\xi ^2 (3 \lambda_2+\lambda_3)\nonumber\\
&\mp& 2 \nu_\xi\textit{F})+(3 \lambda_2+\lambda_3) (16 \nu \xi ^3 (3 \lambda_2+\lambda_3)\nonumber\\
&+&\nu_\phi ^2 (4 \nu_\xi  (2 \lambda_4+\lambda_5)+\mu_1))\nonumber\\
&+&6 \mu_2^2 \nu_\xi) ,\\
\Delta V_0({C_{1(2)}B})&\equiv&V_0(C_{1(2)})-V_0(B) \nonumber\\
&=& -\frac{3}{256 \nu_\xi  (3 \lambda_2+\lambda_3)^3}( \textit{F} \mp 3 \mu_2)^2 (\mu_2 (24 \nu_\xi ^2 (3 \lambda_2+\lambda_3)\nonumber\\
&\mp& 2 \nu_\xi\textit{F})+(3 \lambda_2+\lambda_3) (16 \nu_\xi ^3 (3 \lambda_2+\lambda_3)\nonumber\\
&+&\nu_\phi ^2 (4 \nu_\xi  (2 \lambda_4+\lambda_5)+\mu_1))\nonumber\\
&+&6 \mu_2^2 \nu_\xi)+\lambda_1 \nu_\phi^4 + 3 \nu_\xi^3 (\mu_2 + (3\lambda_2+\lambda_3)\nu_\xi)\nonumber\\
&+&\frac{3}{8} \nu_\xi(\mu_1+4(2 \lambda_4+\lambda_5)\nu_\xi)\nu_\phi^2 ,\\
\Delta V_0({C_1C_2})&\equiv& V_0(C_1)-V_0(C_2) \nonumber\\
&=& \frac{3\mu_2 F^{3}}{16(3\lambda_2+\lambda_3)^3}\;.
\end{eqnarray}\label{eq:potential}

 The $\Delta V_0({C_1C_2})$ determines the detailed phase transition patterns in the two-step phase transition scenario, as will be explored latter.

\subsection{Phase transition dynamics}

\begin{figure}[!tb]
\begin{center}
\includegraphics[width=0.3\textwidth]{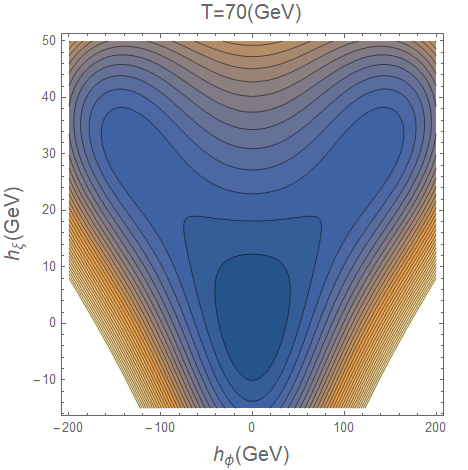}
\includegraphics[width=0.3\textwidth]{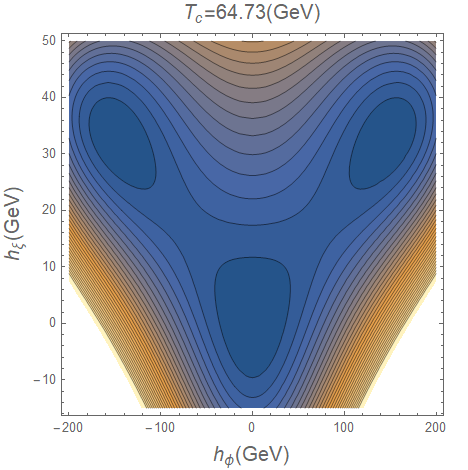}
\includegraphics[width=0.3\textwidth]{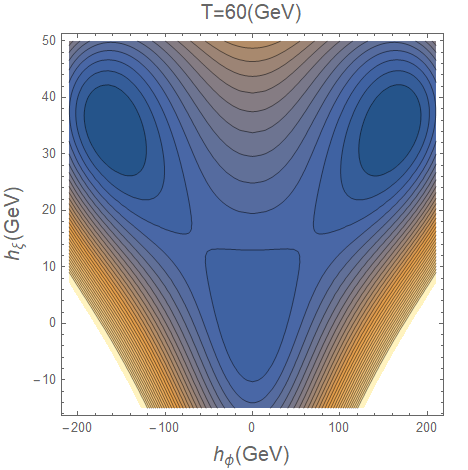}
\end{center}
\caption{The contours of $V_{T}$ in $h_{\phi}$ - $h_{\xi}$ plane, with the parameters being: $\lambda_1 = 0.035$, $\lambda_2 = 0.574$, $\lambda_3 = -0.547$, $\lambda_4 = 0.798 $, $\lambda_5 = 1.908$, $\theta_H = 0.273$, $\mu_1=-360.373$ GeV, $\mu_2=-47.377$ GeV.}\label{fig:1stepoperates}
\end{figure}

With the zero temperature scalar potential at hand, the phase transition dynamics can be estimated with the finite temperature potential using the gauge invariant approach~\cite{Patel:2011th},
\begin{align}
V_{T}=V_{0}+\frac{1}{2}c_{\phi} T^2 h_{\phi}^2 +\frac{1}{2} c_{\xi} T^2 h_{\xi}^2+\frac{1}{2} c_{\chi} T^2 h_{\chi}^2 \;,\label{eq:fP}
\end{align}
with 
$V_{0}$ being given in~\autoref{eq:Vtree}, and the finite temperature corrections being 
\begin{align}
\label{equ:cphichi}
c_{\phi}&=\frac{3 g^2 }{16}+\frac{g'^2 }{16}+ 2 \lambda_1+\frac{3 \lambda _4 }{2}+\frac{1}{4}  y_t^2 \sec ^2\theta_H\;,\nonumber\\
c_{\xi}&=\frac{g^2}{2}+\frac{11 \lambda _2 }{3}+\frac{7 \lambda_3 }{3}+\frac{2 \lambda_4 }{3}\;,\nonumber\\
c_{\chi}&=\frac{g^2 }{2}+\frac{g'^2 }{4}+\frac{11\lambda _2 }{3}+\frac{7 \lambda_3 }{3}+\frac{2 \lambda _4 }{3}\;.
\end{align}
For the safety of custodial symmetry, we assume $h_\chi=\sqrt{2}h_\xi$ for phase transition studies.  
In the one-step phase transition situation, the phase transition occurs through the path of $A\to B$ directly with 
$h_{\phi,\xi}^B$ locating around $v_{\phi,\xi}$ at finite temperature $T_C$.
The phase transition may occur after the temperature drops bellow the $T_C$. In~\autoref{fig:1stepoperates}, we illustrate the one-step phase transition process as the temperature drops. The global minimum of the finite temperature potential $V_T$ changes from the $A$ to the $B$, through which one obtains the EW symmetry breaking minimum. 

As for the two-step phase transition scenario, the condition for the first step of the two-step phase transition ($A~\rm{point} \to C~\rm{point}$) can be written as:
\begin{align}
&V_T(0,0,{T_{1C}}) = V_T(0,h_{\xi}^{1C},{T_{1C}}) \;,
&\left.\frac{dV_T(h_\phi,h_\xi,T_{1C})}{dh_\xi}\right|_{h_\phi = 0,h_\xi = h_{\xi}^{1C}} = 0\;.
\end{align}
The critical parameters $h_{\xi}^{1C}$ and $T_{1C}$ are calculated as,
\begin{align}
&h_{\xi}^{1C} = - \frac{\mu_2}{3 \lambda_2+\lambda_3}\;,
&T_{1C} = \frac{\sqrt{-3 m_2^2 (3 \lambda_2 + \lambda_3) + 6 \mu_2^2}}{\sqrt{(c_\xi + 2 c_\chi)(3 \lambda_2 + \lambda_3)}} \;.
\end{align}
For the second-step of the two-step phase transition to occur, the following degeneracy conditions at the critical temperature are necessary,
\begin{align}
&V_T(0,h_{\xi}^C,T_C) = V(h_{\phi}^B,h_{\xi}^B,T_C) \;,\nonumber\\
&\left.\frac{dV_T(h_\phi ,h_\xi ,T_C)}{dh_\phi}\right|_{{h_\phi } = h_{\phi}^B,{h_\xi } = h_{\xi}^B} = 0\;,
\left.\frac{dV_T(h_\phi ,h_\xi ,T_C)}{dh_\xi }\right|_{{h_\phi } = h_{\phi}^B,{h_\xi } = h_{\xi}^B} = 0 \;,
\left.\frac{dV_T(0,h_\xi ,T_C)}{dh_\xi }\right|_{{h_\xi } = h_{\xi}^C}  = 0 \; ,
\end{align}
through which the critical temperature and critical field value can be obtained. For the two-step case, 
using the determinant of the Hessian matrix (at both the zero temperature and the finite temperature) to ensure the two degenerate vacua occur, the
following conditions need to be satisfied:  $ M_{3(5,6)}P_{3(5,6)} - {N_{3(5,6)}^2} > 0,M_{3(5,6)} > 0 $, with
\begin{align}\label{treehess}
&\frac{{{d^2}V_0({h_\phi },{h_\xi })}}{{dh_\phi ^2}}{|_{{h_\phi } = v_{\phi},{h_\xi} = v_{\xi}}} \equiv M_3\;, \frac{{{d^2}V_0({h_\phi },{h_\xi })}}{{d{h_\phi }d{h_\xi }}}{|_{{h_\phi} = v_{\phi},{h_\xi} = v_{\xi}}} \equiv N_{3}\;, \nonumber\\
&\frac{{{d^2}V_0({h_\phi },{h_\xi })}}{{dh_\xi ^2}}{|_{{h_\phi} = v_{\phi},{h_\xi}= v_{\xi}}} \equiv P_{3}\;.\\
&\frac{{{d^2}V_T({h_\phi },{h_\xi },{T_{2C}})}}{{dh_\phi ^2}}{|_{{h_\phi } = h_{\phi}^B,{h_\xi} = h_{\xi}^B}} \equiv M_5\;, \frac{{{d^2}V_T({h_\phi },{h_\xi },{T_{2C}})}}{{d{h_\phi }d{h_\xi }}}{|_{{h_\phi} = h_{\phi}^B,{h_\xi} = h_{\xi}^B}} \equiv N_{5}\;, \nonumber\\
&\frac{{{d^2}V_T({h_\phi },{h_\xi },{T_{2C}})}}{{dh_\xi ^2}}{|_{{h_\phi} = h_{\phi}^B,{h_\xi}= h_{\xi}^B}} \equiv P_{5}\;,\\
&\frac{{{d^2}V_T(h_\phi,{h_\xi },{T_{2C}})}}{{dh_\phi ^2}}{|_{{h_\phi} = 0,{h_\xi} = h_{\xi}^{2C}}} \equiv M_6\;, 
\frac{{{d^2}V_T(h_\phi,{h_\xi },{T_{2C}})}}{{dh_\phi dh_\xi}}{|_{{h_\phi} = 0,{h_\xi} = h_{\xi}^{2C}}} \equiv N_6\;,\nonumber\\
&\frac{{{d^2}V_T(h_\phi,{h_\xi },{T_{2C}})}}{{dh_\xi ^2}}{|_{{h_\phi} = 0,{h_\xi} = h_{\xi}^{2C}}} \equiv P_6\;.
\end{align}

Here, at finite temperature $T_{2C}$, the $h_{\phi,\xi}^B$ locates around $v_{\phi,\xi}$, and the $h_\xi^{2C}$ locates around $h_{\xi_{C_1(_2)}}$ as given in~\autoref{equ:eqvac}.
That the temperature of the first step phase transition is higher than the second one, i.e., $T_{1C} > T_{2C}$, is also used to select the SFOEWPT points for the two step phase transition scenario.

\begin{figure}[!tb]
\begin{center}
\includegraphics[width=0.24\textwidth]{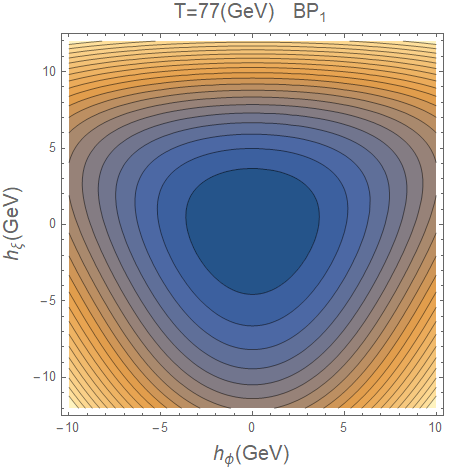}
\includegraphics[width=0.24\textwidth]{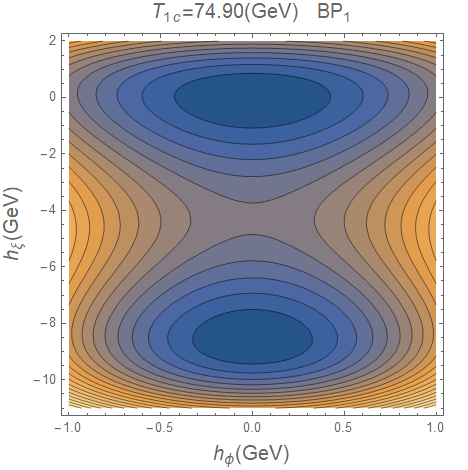}
\includegraphics[width=0.24\textwidth]{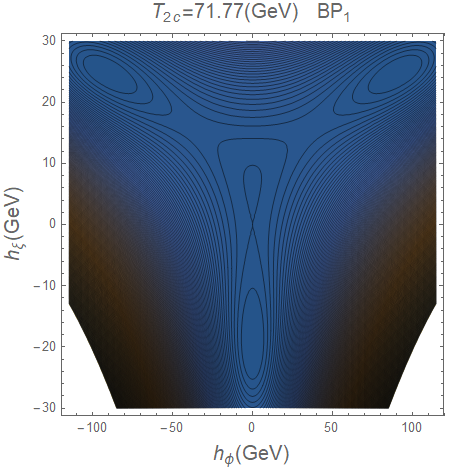}
\includegraphics[width=0.24\textwidth]{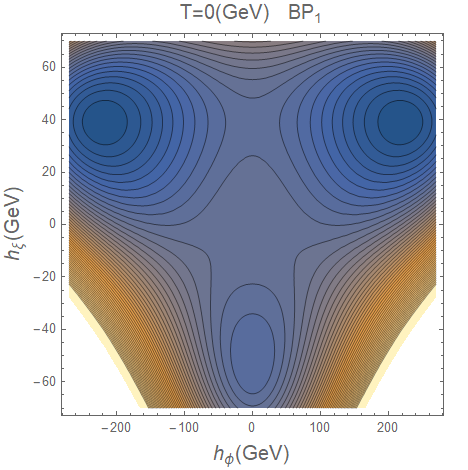}
\includegraphics[width=0.24\textwidth]{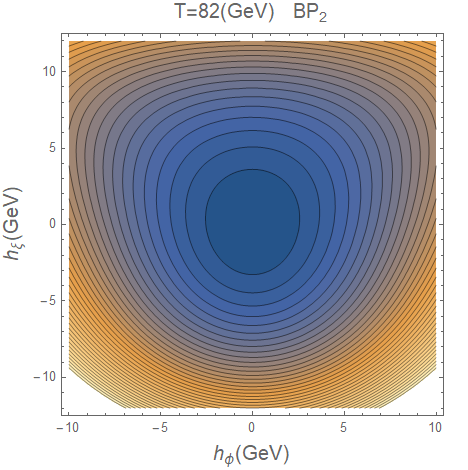}
\includegraphics[width=0.24\textwidth]{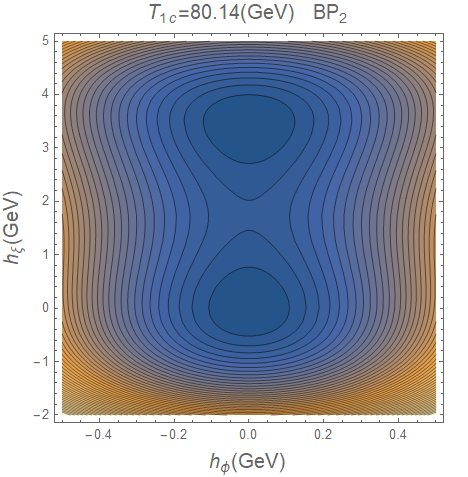}
\includegraphics[width=0.24\textwidth]{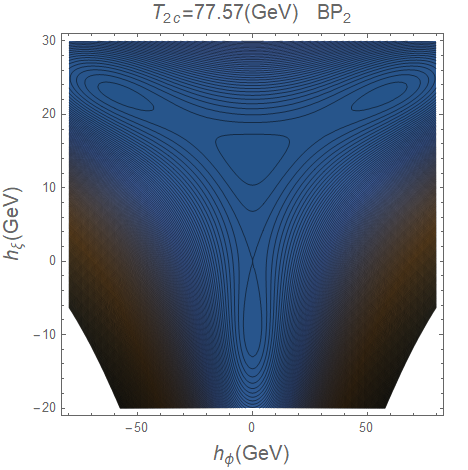}
\includegraphics[width=0.24\textwidth]{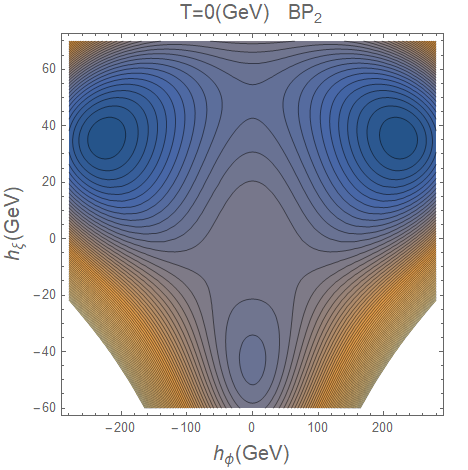}
\end{center}
\caption{The evolution of the vacuum structure as temperature drops. 
 The two-step SFOEWPT point ($A\to C_2\to B$) with the initial phase of the second step locating around $C_2$ for $BP_1$.  
 The two-step SFOEWPT point with the phase transition pattern of $A\to C_1\to B$ and the initial phase of the second step locating around $C_1$ for $BP_2$.
 The parameters for $BP_1$ are: $\lambda_1=0.040$,$\lambda_2=0.598$, $\lambda_3=-0.113$, $\lambda_4=0.425$, $\lambda_5=0.264$, $\theta=0.467$, $\mu_1=-175.619$ GeV, $\mu_2=14.416$ GeV.  The parameters for $BP_2$ are:  $\lambda_1=0.039$,$\lambda_2=0.692$, $\lambda_3=-0.311$, $\lambda_4=0.479$, $\lambda_5=0.457$, $\theta=0.423$, $\mu_1=-181.111$ GeV, $\mu_2=-6.144$ GeV.}\label{fig:2stepbp12}
\end{figure}

At zero temperature, the $C_{1,2}$ can be a local minimum or saddle point with the global vacuum being located at $B$. 
We show in~\autoref{fig:2stepbp12} how the phase transitions occur as the temperature drops in two-step case.
During these two-step phase transition processes, the global vacuum of the finite temperature potential $V_T$ changes from the $A$ to the $C_{1,2}$ at the first-step after $T<T_{1C}$. Then, the phase changes from the vacuum locating around $C_{1,2}$ to the one around $B$, i.e., the EW symmetry breaking vacuum, after $T<T_{2C}$. 

The above procedure is firstly used to obtain the critical phase transition order parameters and the rough phase transition pattern (either one-step or two-step). Subsequently, we use {\tt CosmoTransitons}~\cite{Wainwright:2011kj} to obtain the phase transition order parameters at the bubble nucleation temperature, which might be slightly different with the patterns from the above approach. This is mainly due to the reason that in some cases, although we can obtain the phase transition with the above approach, the improper barrier between the two minima at the bubble nucleation temperature couldn't fulfill the condition of bubble nucleation, see~\autoref{eq:bn}. This is especially important for the two-step cases where the pattern would be changed from $A\to C_{1,2}\to B$ to $A\to B$. Hence, when we present the results, the term ``two-step'' refers to the points obtained by the above approach, and the terms ``bubble one-step'' and ``bubble two-step'' refer to the points that can trigger one-step bubble nucleation and two-step bubble nucleation respectively after we check the bubble nucleation using {\tt CosmoTransitions}.


\section{The SFOEWPT and Higgs phenomenology}
\label{sec:SFOEWPT}

The vacuum structure at zero temperature, such as the potential barrier, is crucial for both one-step and two-step SFOEWPT. The desired vacuum structure for the SFOEWPT reveals the Higgs potential shape with triple and quartic Higgs couplings deviations from the SM case. The typical vacuum structure for one-step and two-step SFOEWPT being explored in the last section can have distinct Higgs phenomenological predictions. 

In the GM model, both the two extra triplets contribute to the EWSB and the gauge bosons get masses also from the triplet VEV, $v_{\xi,\chi}$. The Electroweak charge of the triplets leads to the $HVV$ couplings deviating from the singlet case by one extra factor $\sin\theta_H$, which parameterizes the contribution of the $v_{\xi,\chi}$ to the Higgs VEV, $\sin\theta_H=2\sqrt{2}\nu_\xi/\sqrt{(8v_\xi^2+v_\phi^2)}$~\cite{Hartling:2014zca}.
For the GM model, due to the isospin triplet contribute to the EWSB, 
the phase transition strength is defined as~\cite{Zhou:2018zli},  
\begin{align}
v^{GM}/T&\equiv \frac{\sqrt{v_\phi^2(T)+8v_\xi^2(T)}}{T}=\frac{v_{\phi}(T)\cos\theta_H(T)^{-1}}{T}\;,\nonumber\\
\cos\theta_H(T) &\equiv\frac{v_{\phi}(T)}{\sqrt{v_\phi^2(T)+8v_\xi^2(T)}}\;,
\end{align}
at the critical temperature where phase transition occurs. Since we are working in the scenario where the zero temperature vacuum structure is crucial for the SFOEWPT, the $\theta_H(T)$ here would be highly related with the $\theta_H$ (see our previous studies in Ref~\cite{Zhou:2018zli} for detail). As will be explored latter, one can expect that the one-step and two-step SFOEWPT valid regions are highly restricted by collider searches. 

To study the collider phenomenology of new physics models, one needs to work in the physical basis. 
In terms of physical field basis after taking into account of the rotation matrix (with the angle $\alpha$) among classical fields and Higgs fields $h,H$, the interaction strength between the SM-like Higgs and SM particles are:
\begin{align}
&g_{hf\bar{f}}=\cos\alpha/\cos\theta_H g^{SM}_{hf\bar{f}}\,,~g_{hVV}=(\cos\alpha\cos\theta_H-\sqrt{\frac{8}{3}}\sin\alpha\sin\theta_H) g^{SM}_{hf\bar{f}} \;,\nonumber\\
&
g_{H f\bar{f}}=\sin\alpha/\cos\theta_H g^{SM}_{hf\bar{f}}\, ,~ g_{H VV}=(\sin\alpha\cos\theta_H+\sqrt{\frac{8}{3}}\cos\alpha\sin\theta_H )g^{SM}_{hVV}\;.
\end{align}
Currently, the angle $\theta_H$ is severely bounded by the same-sign WW boson channel search at 13 TeV LHC~\cite{Sirunyan:2017ret}.
Future hadron and lepton colliders would further restrict the magnitude of $\alpha$ and $\theta_H$,
which means that the possibility to reach SFOEWPT would be bounded to the parameter spaces with small $\sin\theta_H$ and small $\alpha$. 
For the case of small $\theta_H$ limit 
where the EWSB contribution from the triplet is negligible, 
one will also have 
\begin{align}
\alpha\approx-\sqrt{\frac{3}{2}} \theta_H+\frac{\theta_H^2 (8 \sqrt{3} \lambda_1  -2 \sqrt{3} \lambda_4  -\sqrt{3} \lambda_5  )\nu}{\mu_1}\;.
\end{align}
This means, for a small $\theta_H$ one usually have a small $\alpha$, and the sign of $\alpha$ is determined by the coupling combinations of $8 \sqrt{3} \lambda_1  -2 \sqrt{3} \lambda_4  -\sqrt{3} \lambda_5 $ and $\mu_1$. In the scenario with small $\alpha$ and small $\theta_H$, the $g_{hf\bar{f},hVV}$ close to the SM case, $g_{Hf\bar{f},HVV}$ are suppressed.

The scalar potential of~\autoref{eq:Vtree} in the Higgs basis of $h$ and $H$ can be written as,
\begin{eqnarray}
V^{GM}_{phy}&=&
\frac{1}{2} m_1^2 (h \cos\alpha+H \sin\alpha)^2 +\frac{1}{2} m_2^2 (H \cos\alpha-h \sin\alpha)^2\nonumber\\
&&+\frac{2}{\sqrt{3}} \mu_2 (H \cos\alpha-h \sin\alpha)^3+\frac{\sqrt{3}}{4} \mu_1 (h \cos\alpha+H \sin\alpha)^2 (H \cos\alpha-h \sin\alpha)\nonumber\\
&&+\lambda_1 (h \cos\alpha+H \sin\alpha)^4+(\lambda_2+\frac{1}{3} \lambda_3)(H \cos\alpha- h \sin\alpha)^4 \nonumber\\
&&+(\lambda_4+\frac{1}{2} \lambda_5)(h \cos\alpha+H \sin\alpha)^2 (H\cos\alpha-h \sin\alpha)^2\;,
\end{eqnarray}
for the GM model. 
In the small $\alpha$ limit, the potential $V^{GM}_{phy}$ reduces to
\begin{eqnarray}
V^{GM}_{\alpha}&=&\frac{1}{2}m_1^2 h^2+\lambda_1 h^4
+\frac{1}{2}m_2^2 H^2+\frac{2\sqrt{3}}{3}\mu_2 H^3+\frac{1}{3}(3\lambda_2+\lambda_3)H^4+\frac{\sqrt{3}}{4}\mu_1 h^2 H\nonumber\\
&&+\frac{1}{2}(2\lambda_4+\lambda_5) h^2 H^2 +\alpha h(-\frac{\sqrt{3}\mu_1}{4} h^2+(m_1^2-m_2^2)H+\frac{\sqrt{3}}{2}(\mu_1-4\mu_2)H^2\nonumber\\
&&+(-4\lambda_2-\frac{4\lambda_3}{3}+2\lambda_4+\lambda_5)H^3+(4\lambda_1-2\lambda_4-\lambda_5)h^2H)+ \mathcal{O}(\alpha^2)\;.\label{eq:GMalp}
\end{eqnarray}
After EWSB, $h$ and $H$ get VEVs,  
\begin{eqnarray}
&&v_h^{GM} = v \cos(\alpha)\cos(\theta_H)-\frac{1}{2} \sqrt{\frac{3}{2}} v \sin(\alpha) \sin(\theta_H) \;,\nonumber\\
&&v_H ^{GM}= v \sin(\alpha)\cos(\theta_H)+\frac{1}{2} \sqrt{\frac{3}{2}} v \cos(\alpha) \sin(\theta_H)\;,
\end{eqnarray} 
with $v=v_{SM}\equiv 246$ GeV, and $v_\xi=v\sin\theta_H/2\sqrt{2}$.
Suppose $h$ is the SM-like Higgs, one has both the Higgs cubic and quartic couplings being modified comparing with the SM case. This can be parameterized as:
\begin{eqnarray}
  \Delta \mathcal{L} = - \frac{1}{2}\frac{m_{h}^2}{v} (1 + \delta \kappa_3) h^3 - 
  \frac{1}{8} \frac{m_{h}^2}{v^2} (1 + \delta \kappa_4) h^4 ,
\end{eqnarray}
The cubic Higgs couplings are crucial for the vacuum structure and therefore the phase transition dynamics, as well as the Higgs pair production at hadron and lepton colliders. Consequently, the Higgs pair searches could be powerful to probe the parameter space of the SFOEWPT. 
In the small $\alpha$ ($\theta_H$) limit, we have 
\begin{eqnarray}
&&\delta\kappa^{GM}_3 = -\alpha\frac{\sqrt{3}\mu_1 v}{2m_h^2}+\frac{\alpha v^2 (4\alpha-\sqrt{6}\theta_H)(2\lambda_4+\lambda_5)}{2m_h^2}-\frac{(3\alpha^2+\theta_H^2)}{2}+\mathcal{O}(\alpha^3,\theta_H^3)\;,\\
&&\delta\kappa^{GM}_4=-2\alpha^2\left(1-\frac{2(2\lambda_4+\lambda_5)v^2}{m_h^2}\right)+\mathcal{O}(\alpha^3)\;.
\end{eqnarray}

As a comparison, we also list the case for xSM model which has no extra EWSB contribution.
In xSM case, we also define a mixing angle $\alpha$ between the SM Higgs ($h$) and extra scalar ($s$). At the zero temperature, it is defined as~\cite{Profumo:2014opa}
 \begin{align}
 \label{equ:xSMalpha}
 \sin2\alpha=\frac{(a_1+2 a_2 v_s)v_h}{(m_{h}^2-m_{H}^2)}\;,
 \end{align} 
with $v_h=246$ GeV. For $-1\leq \sin2\alpha\leq 1$, which sets a bound on the VEV fraction of the SM Higgs ($h$), 
\begin{align}
m_{h}^2- m_{H}^2\leq (a_1+2 a_2 v_s)v_h \leq m_{H}^2- m_{h}^2\;.
\end{align} 
The phase transition occurs in the subspace of the two scalar fields, with the phase transition strength being 
\begin{align}
v^{\rm xSM}/T&\equiv\frac{v_h(T)}{T}= \frac{\sqrt{v_h^2(T)+v_s^2(T)}\cos\theta(T)}{T}\;,\nonumber\\
\cos\theta(T)&\equiv\frac{v_h(T)}{\sqrt{v_h^2(T)+v_s^2(T)}}\;.
\end{align}
The sphaleron process is quenched when  $v_h(T)/T>1$ at the critical temperature.  Different from the GM model, the singlet VEV ($v_s$) does not contribute to the mass of gauge boson due to the Electroweak charge of the singlet, hence it does not contribute to the phase transition strength.

The collider search of the Higgs phenomenology is performed in the basis of the SM Higgs and one extra heavy Higgs. The current LHC Higgs data and theoretical constraints require small $\alpha$ which parameterizes the mixing between the SM-like Higgs and the extra CP-even heavy Higgs.
In this model, all the couplings are rescaled by $\alpha$ based on the SM as: $g_{h xx}=\cos\alpha g^{SM}_{h xx}, g_{H xx}=-\sin\alpha g^{SM}_{h xx}$. Therefore, no direct bound on the angle $\theta(T=0)$ from the Higgs data since the parameter does not enter Higgs couplings. The VEV of the extra scalar ($v_s$) would be more free than that in GM model.

\begin{figure}[!tbp]
\begin{center}
\includegraphics[width=0.395\textwidth]{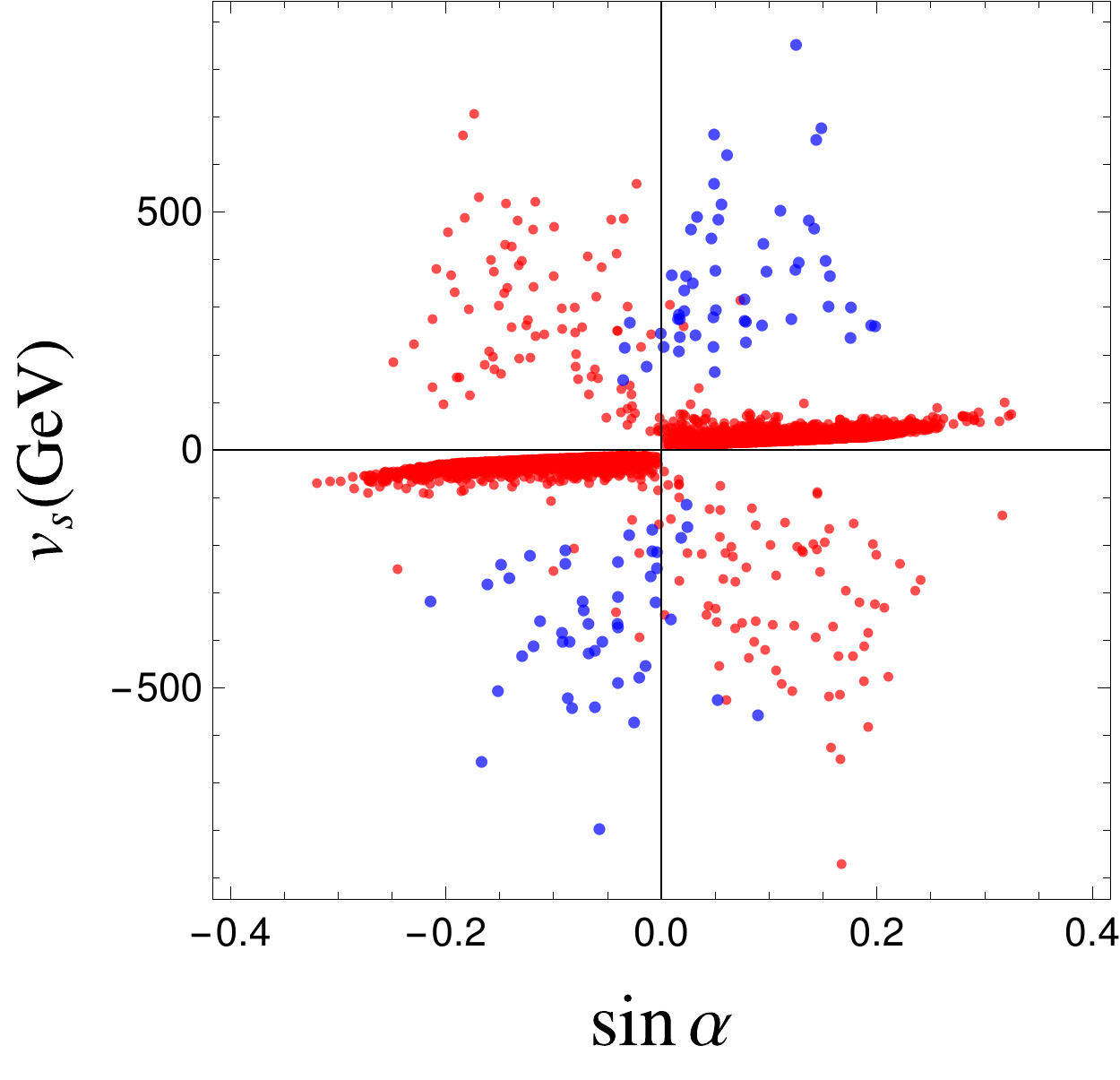}
\includegraphics[width=0.4\textwidth]{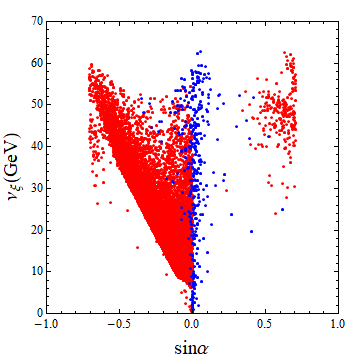}
\end{center}
\caption{The $v_C/T_C>1$ viable points (both one-step (red) and two-step (blue)) in the $\sin\alpha$-$v_{s,\xi}$ plane for the xSM (left) and the GM (right) model.}
\label{fig:thalpha}
\end{figure}

For the xSM model, the potential in the basis of Higgs fields is given by,
\begin{eqnarray}
V^{\rm xSM}_{phy}&=&\frac{1}{12} (3 a_1 (h \cos\alpha -H \sin\alpha)^2 (h \sin\alpha +H \cos\alpha)+3 a_2 (h \cos\alpha -H \sin\alpha )^2 \nonumber\\
&&\times (h \sin\alpha+H \cos\alpha )^2
+6 b_2 (h \sin\alpha+H \cos\alpha )^2+4 b_3 (h   \sin\alpha +H \cos\alpha )^3\nonumber\\
&&+3 b_4 (h \sin\alpha+H \cos\alpha )^4
+3 \lambda  (h \cos \alpha-H \sin\alpha )^4-6 \mu ^2 (h \cos\alpha -H  \sin\alpha )^2)\;.
\end{eqnarray}
Going to the alignment case, one have,
\begin{eqnarray}
V^{\rm xSM}_{\alpha}&=&
\frac{1}{12} (3 a_1 h^2 H+3 a_2 h^2 H^2+6 b_2 H^2+4 b_3 H^3+3 b_4 H^4+3 h^4 \lambda -6 h^2 \mu ^2)\nonumber\\
&&+\frac{1}{4} h \alpha  (a_1 (h^2-2 H^2)+2 H (a_2 (h^2-H^2)+2 b_2+2 (H (b_3+b_4 H)-\lambda h^2 +\mu ^2)))\nonumber\\
&&+ \mathcal{O}(\alpha ^2)\;.
\label{eq:xSMalp}
\end{eqnarray}
After EWSB, 
one have for the two physical fields $h$ and $H$:
\begin{align}
v_h^{\rm xSM} = v_h\cos(\alpha)+ v_s \sin(\alpha)\;,\nonumber\\
v_H^{\rm xSM} = v_s \cos(\alpha) - v_h \sin(\alpha)\;.
\end{align}
In the xSM, the deviation of the cubic and quartic couplings for small $\alpha$ are given by~\cite{Alves:2018jsw}:
\begin{eqnarray}
&& \delta \kappa^{\rm xSM}_3 = \alpha^2 \left[-\frac{3}{2} + \frac{2 m_{H}^2 -2 b_3 v_s - 4 b_4 v_s^2}{m_{h}^2}\right] + \mathcal{O}(\alpha^3)\;, \nonumber \\
&& \delta \kappa^{\rm xSM}_4 = \alpha^2 \left[-3 + \frac{5 m_{H}^2 - 4 b_3 v_s - 8 b_4 v_s^2}{m_{h}^2}\right] + \mathcal{O}(\alpha^3)  \;. 
\end{eqnarray}

Due to the extra EWSB contribution, in the GM model, one has an additional angle $\theta_H$ to parameterize the Higgs couplings. Therefore one has the different distributions of SFOEWPT points in the $v_\xi$-$\alpha$ plane, which builds the bridge between the SFOEWPT and the Higgs phenomenology.
\autoref{fig:thalpha} shows the one-step (red) and two-step (blue) SFOEWPT valid points in the xSM (left) and GM (right) model. For the xSM case, the one-step (two-step) SFOEWPT points concentrate in the small (large) $v_s$ regions. While in the GM model, the $v_\xi$ is much smaller than the $v_s$ in xSM and the possibility to reach a SFOEWPT drops with the decrease of the $v_\xi$ and $|\sin\alpha|$. 

\begin{figure}[!tbp]
\begin{center}
\includegraphics[width=\textwidth]{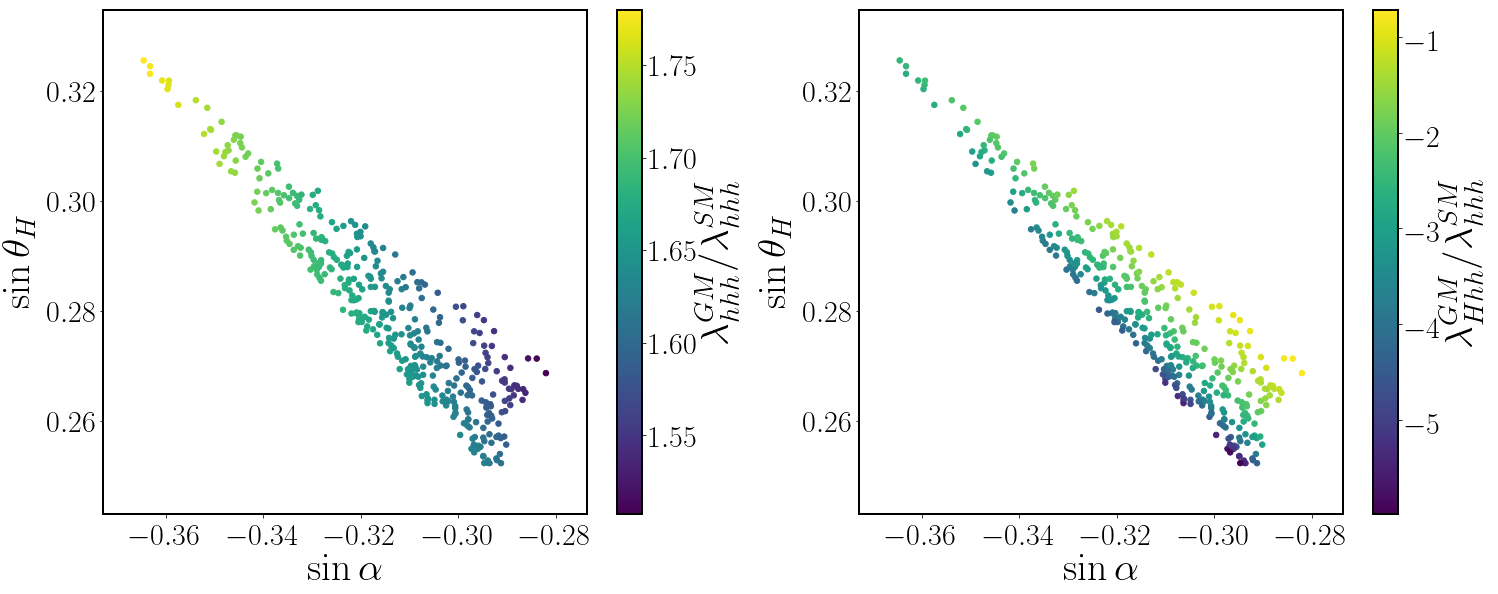}
\end{center}
\caption{
The ratio of $\lambda_{hhh}^{GM}/\lambda_{hhh}^{SM}$ (left) and $\lambda_{Hhh}^{GM}/\lambda_{hhh}^{SM}$ (right) in $\sin\alpha$-$\sin\theta_H$ plane in \hp{}.
}
\label{fig:h5plane_3h}
\end{figure}

In the parameter spaces which allow the SFOEWPT, one has the deviations of the cubic and quartic Higgs couplings, which characterize the Higgs potential shape that are crucial for the realization of the EWSB mechanism. 
We investigate more details in two particular benchmarks of the GM model: the \hp{} which is developed by the LHC Higgs Cross Section Working Group for fiveplet searches~\cite{deFlorian:2016spz} and the \lm{} studied in~\cite{Logan:2018wtm} for lower mass region. Both benchmark scenarios will lead to interesting searches at the collider.

\begin{figure}[!tbp]
\begin{center}
\includegraphics[width=\textwidth]{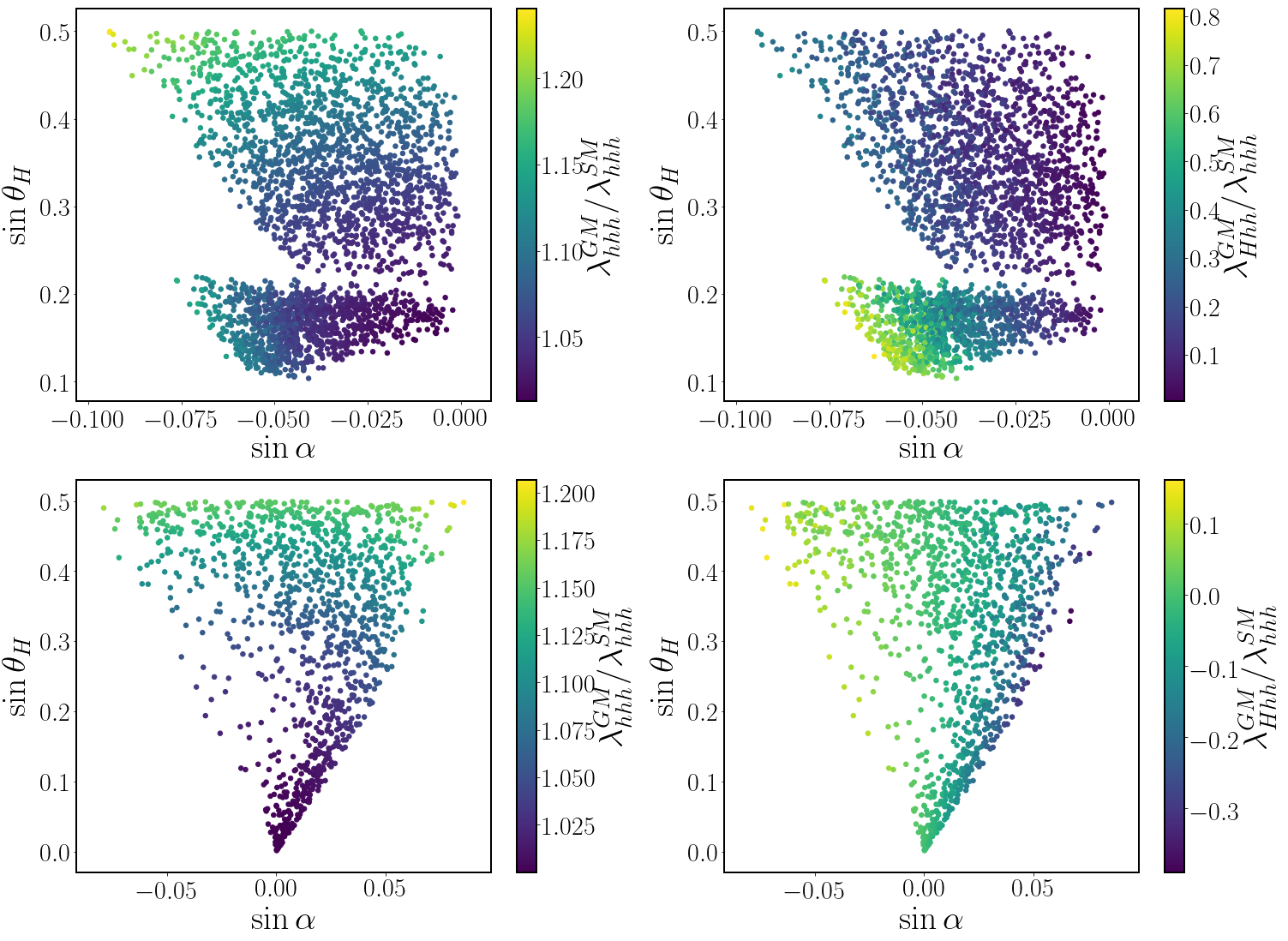}
\end{center}
\caption{
The ratio of $\lambda_{hhh}^{GM}/\lambda_{hhh}^{SM}$ (left panels) and $\lambda_{Hhh}^{GM}/\lambda_{hhh}^{SM}$ (right panels) in $\sin\alpha$-$\sin\theta_H$ plane for both one-step points (upper panels) and two-step points (lower panels) in \lm. 
}
\label{fig:low3h}
\end{figure}

Our previous study in~\cite{Zhou:2018zli} shows that, 
only one-step SFOEWPT is valid in \hp{}, while the \lm{} can provide both one-step and two-step SFOEWPT.
After considering the current LHC search bounds, especially the same-sign W search from CMS~\cite{Sirunyan:2017ret}, we show the triple scalar couplings in~\autoref{fig:h5plane_3h} and~\autoref{fig:low3h} for \hp{} and \lm{} respectively.
In \hp{}, \autoref{fig:h5plane_3h} shows that a larger deviation of the triple Higgs coupling $\lambda_{hhh}^{GM}$ from the SM one occurs with a higher magnitude of $|\alpha|$ and a larger $\theta_H$. While the $\lambda_{Hhh}^{GM}$ is highly enhanced with an extra sign compared with 
$g_{hhh}^{SM}$ which can result in destructive interference for Higgs pair production at the collider.

In \lm{}, one has smaller $\sin\theta_H$ and $m_{h_5}$ in comparison with the \hp{} scenario. 
The CMS same-sign W search severely bounds the $\sin\theta_H$ and thus the mixing angle $\alpha$. Therefore one can expect the triple Higgs coupling and the quartic Higgs couplings are 
all restricted.
In~\autoref{fig:low3h}, we show the triple Higgs couplings in $\sin\alpha$-$\sin\theta_H$ plane for both one (upper panels) and two-step (lower panels) SFOEWPT. 
The SFOEWPT viable points in \lm{} locate in smaller value of $\theta_H$ and $\alpha$. Thus we have lower enhancement in $\lambda_{hhh}^{GM}$ than that in \hp{}. Meanwhile, the $\lambda_{Hhh}^{GM}$ is also much more smaller than that in \hp{}. These will result in different gravitational wave 
production and collider phenomenology as we will study in the following.

A SFOEWPT can also be reached with the help of the dimensional six operator $(H^\dag H)^3$~\cite{Grojean:2004xa,Huang:2015tdv} or new physics that can contribute to such operator. The collider couldn't tell the detailed potential shape (i.e., the tree level potential barrier ) that drives the phase transition. 
In this case, the gravitational wave search for the signal generated by the SFOEWPT would be complementary, since it captures the tunneling process 
manifested in terms of the vacuum bubble nucleations~\cite{Alves:2018oct}.

\section{Gravitational wave searches}
\label{sec:GWTn}

When the temperature of the Universe further cools down after the critical temperature $T_C$ (where one has the degeneracy of the true and the false vacuum), one may have vacuum bubble nucleations, expansions and collisions, and therefore generate GW signals from the SFOEWPT process.   

The bounce configuration of the nucleation bubble (the bounce configuration of the multi-fields that connects the EW broken vacuum ($h$-vacuum, the true vacuum locate around B point) and the false vacuum (the vacuum locate around A or C points)) can be obtained by extremizing
\begin{eqnarray}
S_3(T)=\int 4\pi r^2d r\bigg[\frac{1}{2}\big(\frac{d \phi_b}{dr}\big)^2+V(\phi_b,T)\bigg]\;,
\end{eqnarray}
through solving the equation of motion for $\phi_b$ (it is $h$ and $h_\xi$ for two-step scenarios),
\begin{eqnarray}
\frac{d^2\phi_b}{dr^2}+\frac{2}{r}\frac{d\phi_b}{dr}-\frac{\partial V(\phi_b)}{\partial \phi_b}=0\;,
\end{eqnarray}
with the boundary conditions of 
\begin{eqnarray}
\lim_{r\rightarrow \infty}\phi_b =0\;, \quad \quad \frac{d\phi_b}{d r}|_{r=0}=0\;.
\end{eqnarray}
The phase transition completes at the nucleation temperature when the thermal tunnelling probability for bubble nucleation per horizon volume and per horizon time is of order unity~\cite{Affleck:1980ac,Linde:1981zj,Linde:1980tt}:
\begin{eqnarray}\label{eq:bn}
\Gamma\approx A(T)e^{-S_3/T}\sim 1\;.
\end{eqnarray}

One of the crucial parameter for the gravitational wave is $\alpha$, which is the energy budget of SFOEWPT normalized by the radiative energy, defined as
\begin{eqnarray}
\alpha=\frac{\Delta\rho}{\rho_R}\;,
\end{eqnarray}
where the radiation energy of the bath or the plasma background $\rho_R$ is given by
\begin{eqnarray}
\rho_R=\frac{\pi^2 g_\star T_\star^4}{30}\;,
\end{eqnarray}
The parameter $\Delta\rho$ is the latent heat (vacuum energy density or energy budget of SFOEWPT) from the phase transition to the energy density of the radiation bath
or the plasma background. This is given by the difference of the energy density between the false (here it is $\phi$ vacuum, $\rho(\phi_n,T)$) and the true vacuum (the $h$-vacuum or EW broken vacuum, $\rho(v_n,T)$),
\begin{eqnarray}
\rho(\phi_n,T_n)&=& -V(\phi,T)|_{T=T_n}+T\frac{d\, V(\phi,T)}{d\,T}|_{T=T_n}\;,
\\
\rho(v_n,T_n)&=& -V(h,T)|_{T=T_n}+T\frac{d\, V(h,T)}{d\,T}|_{T=T_n}\;.
\end{eqnarray}
Another crucial parameter $\beta$ characterizes the inverse time duration of the SFOEWPT and thus the GW spectrum peak frequency is defined as
\begin{eqnarray}
\frac{\beta}{H_n}=T\frac{d (S_3(T)/T)}{d T}|_{T=T_n}\; ,
\end{eqnarray}
with $H_n$ being the Hubble constant at the bubble nucleation temperature $T_n$.

\begin{figure}[!tbp]
\begin{center}
\includegraphics[width=\textwidth]{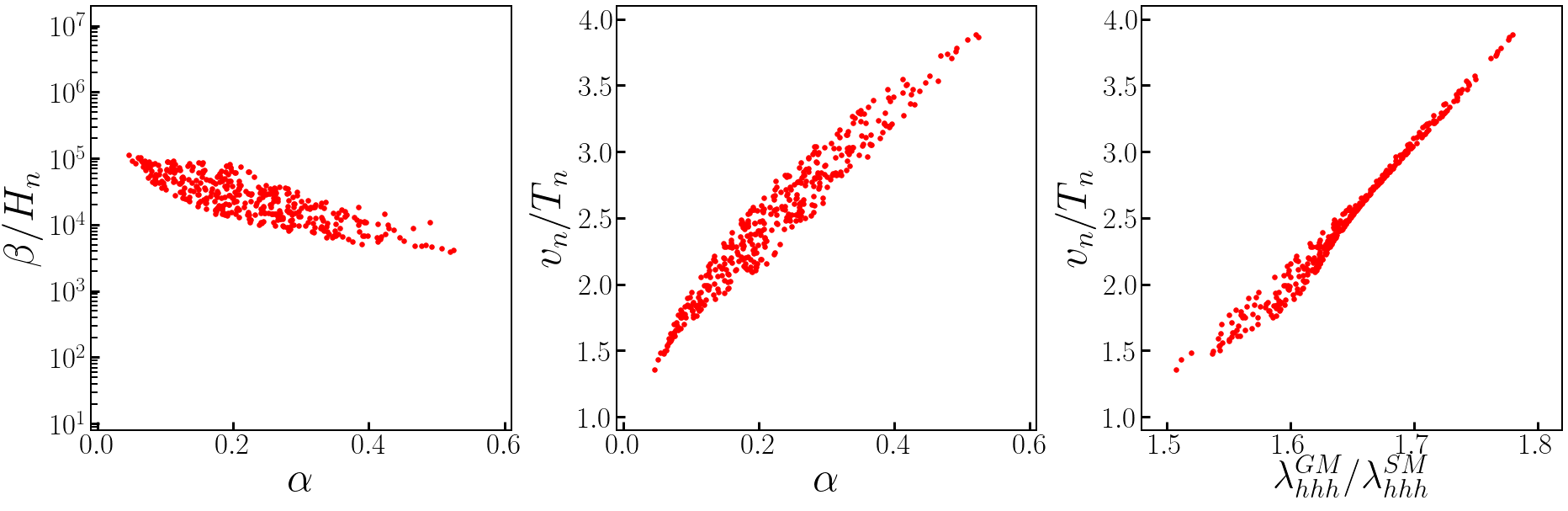}
\end{center}
\caption{
The scanned results in $\alpha$-$\beta/H_n$ (left), $\alpha$-$v_n/T_n$ (middle) and $\lambda_{hhh}^{GM}/\lambda_{hhh}^{SM}$-$v_n/T_n$ (right) planes for \hp{} where we have only one-step (red) points.
}
\label{fig:betalapha_H5plane}
\end{figure}

\begin{figure}[!htp]
\begin{center}
\includegraphics[width=\textwidth]{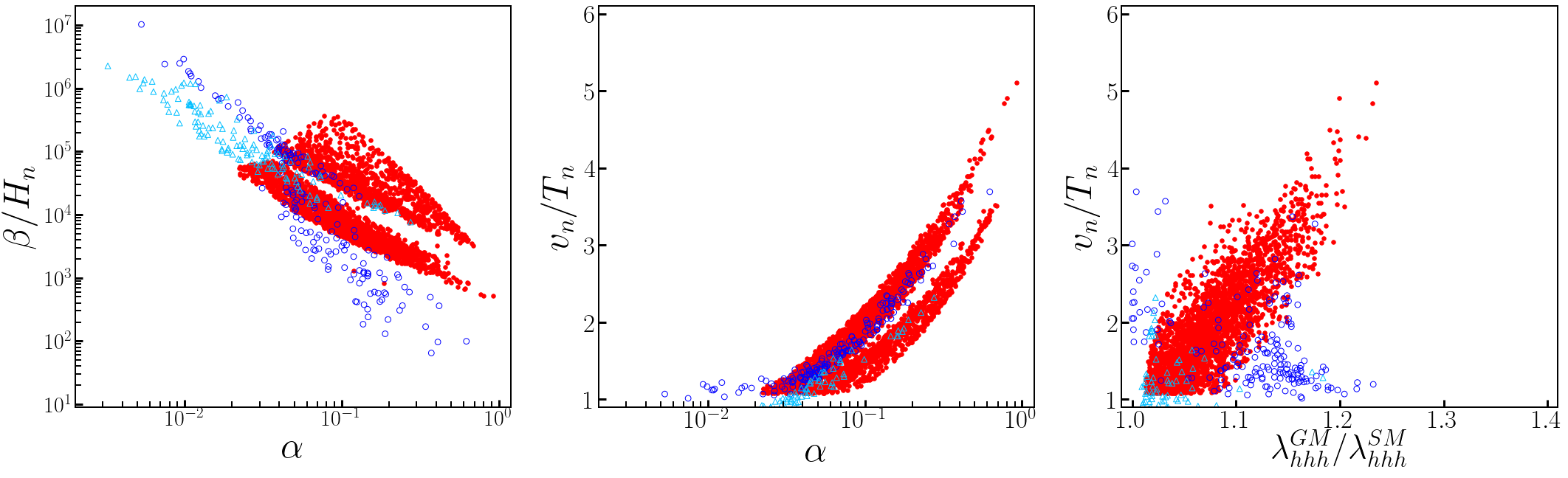}
\end{center}
\caption{
The scanned results in $\alpha$-$\beta/H_n$ (left), $\alpha$-$v_n/T_n$ (middle) and $\lambda_{hhh}^{GM}/\lambda_{hhh}^{SM}$-$v_n/T_n$ (right) planes for \lm{} where we have both one-step (red) and two-step (blue) points. Among the blue points, the circle points represent the case where 
  we can have two-step bubble nucleation (``bubble two-step''), while the triangle points are those where we have only one bubble nucleation (``bubble one-step'').
}
\label{fig:betaalpha_lowmass}
\end{figure}

Considering all the constraints from the LHC, especially the same-sign W bounds from CMS~\cite{Sirunyan:2017ret}, we show the results in~\autoref{fig:betalapha_H5plane} and~\autoref{fig:betaalpha_lowmass} for \hp{} and \lm{} respectively, where the red points represent the one-step scenario and the blue points represent the two-step case. As mentioned before, we separate the two-step points into two groups according to the scan results from {\tt CosmoTransition}. The dark blue circle points represent the case where we have two-step bubble nucleation (``bubble two-step''), while the light blue triangle points are those we have only one bubble nucleation (``bubble one-step''). 

In either \hp{} or \lm{} (one-step and two-step), the $\beta/H_n$ decreases with the increase of $\alpha$. While $\alpha$ is also found to be proportional to the phase transition strength $v_n/T_n$, larger value of $\alpha$ is obtained at larger value of $v_n/T_n$ which results in a relatively larger energy density of the gravitational wave spectrum.
As a comparison, beside that the triple/quartic scalar coupling is smaller in the \lm{}, the \lm{} also provides lower $\beta/H_n$ in the similar range of $\alpha$ than that in \hp{}, which is necessary to produce detectable gravitational wave. In \lm{} where we have both one-step and two-step scenarios, two-step SFOEWPT will have relatively smaller $\beta/H_n$ and larger $v_n/T_n$ for the same $\alpha$ than one-step scenario. 

Here the most stringent constraint comes from the CMS same-sign diboson search~\cite{Sirunyan:2017ret} which, however, doesn't extend to mass ($m_5$) below 200 GeV. This is not relevant for \hp{}, since all those points in~\autoref{fig:betalapha_H5plane} have $m_5>200$ GeV. However, in \lm{}, we can clearly see two parts of points for one-step scenario in~\autoref{fig:betaalpha_lowmass}~\footnote{We do have two parts for two-step as well. However, the points with mass above 200 GeV is minority, we don't discuss them separately.}. Those with lower $\beta/H_n$ or larger $v_n/T_n$, hence resulting in more detectable GW signal, are the points having mass below 200 GeV, which require more dedicated searches at collider.
The triple Higgs couplings are also shown with respect to the SM scenario. The triple Higgs coupling is proportional to phase transition strength $v_n/T_n$ for one-step points in both \hp{} and \lm{}, while there is no clear relation for the two-step points in \lm{}. 

\begin{figure}[!tbp]
\begin{center}
\includegraphics[width=\textwidth]{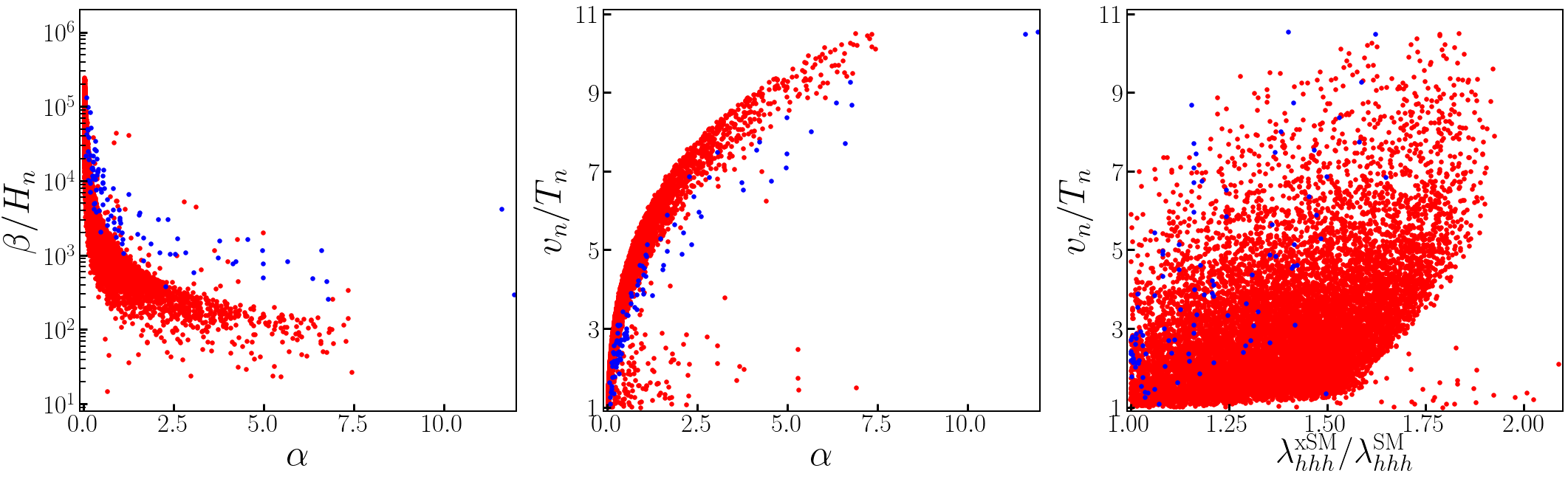}
\end{center}
\caption{
SFOEWPT valid points in $\alpha$-$\beta/H_n$ (left), $\alpha$-$v_n/T_n$ (middle) and $\lambda_{hhh}^{\rm xSM}/\lambda_{hhh}^{SM}$-$v_n/T_n$ (right) planes for xSM. Red and blue points represent one and two-step scenario respectively.
}
\label{fig:xsmGW}
\end{figure}

At last, we comment on the situation in xSM. 
In~\autoref{fig:xsmGW}, we show the relation among $\beta/H_n$, $\alpha$, $v_n/T_n$ and triple scalar couplings in xSM~\footnote{
In this and the following plots for the xSM, we use the same set of data points as used in Ref.~\cite{Alves:2018jsw}.
}.
Since the singlet VEV $v_s$ has no restrictions in general in xSM, the magnitude of $\beta/H_n$ ($v_n/T_n$ and $\alpha$) is relatively much lower (larger) than that in GM model, which results in more detectable GW signal, while the triple scalar couplings has similar enhancement as in GM model.

\begin{figure}[!tbp]
\begin{center}
\includegraphics[width=0.4\textwidth]{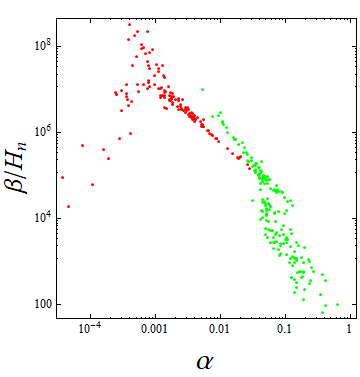}
\includegraphics[width=0.4\textwidth]{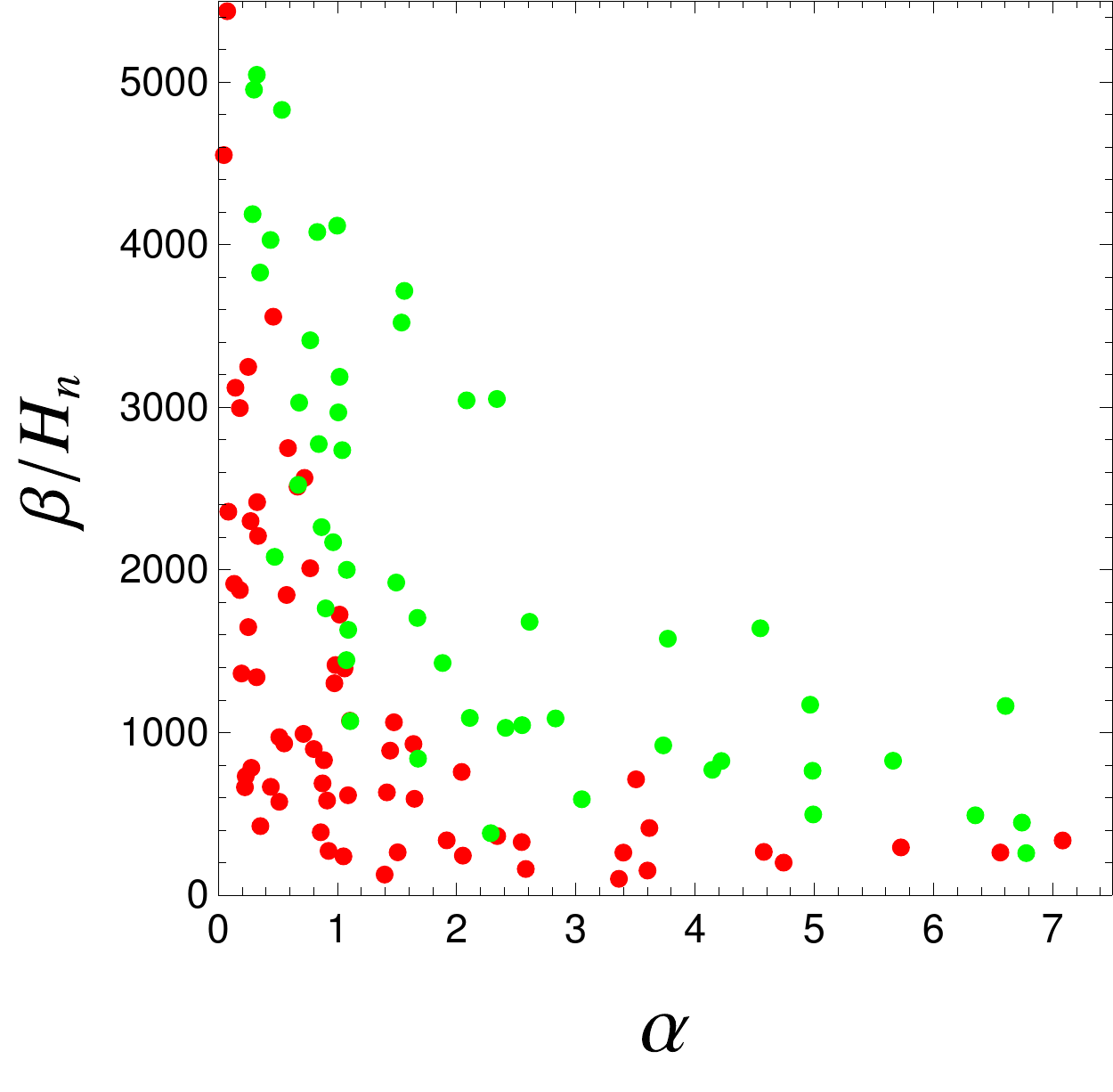}
\caption{The GW signal parameters for the case with extra EWSB (GM) and without EWSB (xSM). The red and green points correspond to the first and the second step of the two-step SFOEWPT. }
\label{fig:GWcol3}
\end{center}
\end{figure}

In~\autoref{fig:GWcol3}, we show the two crucial parameters for the GW signals. The left panel indicates the scenario with EWSB contribution. 
Due to the extra EWSB contribution is subject to severely bounds from the LHC, the extra VEV is small. Which limit the vacuum structure that is important for the phase transition, and thus the first step phase transition can be weakly first order and the GW signal being generated are negligible (with a small $\alpha$ and large $\beta/H_n$), the GW signals in this case mostly dominated by the second-step SFOEWPT which characterize the dynamical EWSB. The right panel is shown here for comparison, one may find that the $\alpha$ and $\beta/H_n$ of the first-step phase transition is mostly smaller than the second-step, which imply a GW signal mostly come from the first-step which does not characterize any symmetry breaking, see Ref.~\cite{Alves:2018jsw} for details.

There are mainly two sources for GW production during the EWPT: the sound waves in the plasma~\cite{Hindmarsh:2013xza,Hindmarsh:2015qta} and 
the magnetohydrodynamic turbulence (MHD)~\cite{Hindmarsh:2013xza,Hindmarsh:2015qta} while the contribution from bubble wall collisions~\cite{Kosowsky:1991ua,Kosowsky:1992rz,Kosowsky:1992vn,Huber:2008hg,Jinno:2016vai,Jinno:2017fby} are now generally
believed to be negligible~\cite{Bodeker:2009qy}. The energy density spectrum from the sound waves can be well expressed by~\cite{Hindmarsh:2015qta}
\begin{eqnarray}
  &&\Omega_{\textrm{sw}}h^{2}=2.65\times10^{-6}\left( \frac{H_{\ast}}{\beta}\right) \left(\frac{\kappa_{v} \alpha}{1+\alpha} \right)^{2} 
\left( \frac{100}{g_{\ast}}\right)^{1/3} 
 v_{w} \left(\frac{f}{f_{\text{sw}}} \right)^{3} \left( \frac{7}{4+3(f/f_{\textrm{sw}})^{2}} \right) ^{7/2} \ .
\label{equ:soundwaves}
\end{eqnarray}
where $H_{\ast}$ is the Hubble parameter at the temperature $T_{\ast}$, at the time when the EWPT finishes;
$v_w$ is the bubble wall velocity; $\alpha$ is the energy released from the EWPT normalized by the total
radiation energy density at $T_{\ast}$ as mentioned above; $g_{\ast}$ is the corresponding relativistic degrees of freedom 
making up the radiation energy density; $\beta$ characterizes roughly the inverse time duration of 
the EWPT. Practically, $T_{\ast}$ is very close to $T_n$ and we use $T_n$ in the following calculations.
Moreover $\kappa_v$ is the fraction of released energy going to the kinetic energy of the plasma, which can 
be calculated given $v_w$~\footnote{
A significant GW production usually needs a very relativistic value of the $v_w$, which however  
is dangerous for baryon asymmetry generation. To deal with this conundrum, we follow Ref.~\cite{No:2011fi,Alves:2018jsw,Alves:2018oct} 
by taking the plasma hydrodynamics into account and distinguish between $v_w$ and the velocity used in baryogenesis calculations.
Therefore a supersonic $v_w$ can be realized while still maintaining a subsonic plasma velocity outside the bubble wall in the wall frame.
} and $\alpha$~\cite{Espinosa:2010hh}.
Finally $f_{\text{sw}}$ is the peak frequency of above energy density spectrum:
 \begin{equation}
f_{\textrm{sw}}=1.9\times10^{-5}\frac{1}{v_{w}}\left(\frac{\beta}{H_{\ast}} \right) \left( \frac{T_{\ast}}{100\textrm{GeV}} \right) \left( \frac{g_{\ast}}{100}\right)^{1/6} \textrm{Hz} ,
\end{equation}
A small fraction of the energy goes to the MHD, whose contribution to the energy density spectrum can also be expressed as~\cite{Caprini:2009yp,Binetruy:2012ze}
\begin{eqnarray}
  &&\Omega_{\textrm{turb}}h^{2}=3.35\times10^{-4}\left( \frac{H_{\ast}}{\beta}\right) \left(\frac{\kappa_{\text{turb}} 
\alpha}{1+\alpha} \right)^{3/2} \left( \frac{100}{g_{\ast}}\right)^{1/3} 
 v_{w}  \frac{(f/f_{\textrm{turb}})^{3}}{[1+(f/f_{\textrm{turb}})]^{11/3}(1+8\pi f/h_{\ast})} , \quad
\label{eq:mhd}
\end{eqnarray}
where $\kappa_{\text{turb}}$ is the fraction of energy going to the MHD and following previous analyses, we take
here $\kappa_{\text{turb}}\approx 0.1 \kappa_{v} $.
Similar to $f_{\text{sw}}$, $f_{\text{turb}}$ is the peak frequency for the spectrum from MHD:
\begin{equation}
f_{\textrm{turb}}=2.7\times10^{-5}\frac{1}{v_{w}}\left(\frac{\beta}{H_{\ast}} \right) \left( \frac{T_{\ast}}{100\textrm{GeV}} \right) \left( \frac{g_{\ast}}{100}\right)^{1/6} \textrm{Hz} .
\end{equation}

\begin{figure}[!htp]
\begin{center}
\includegraphics[width=0.45\textwidth]{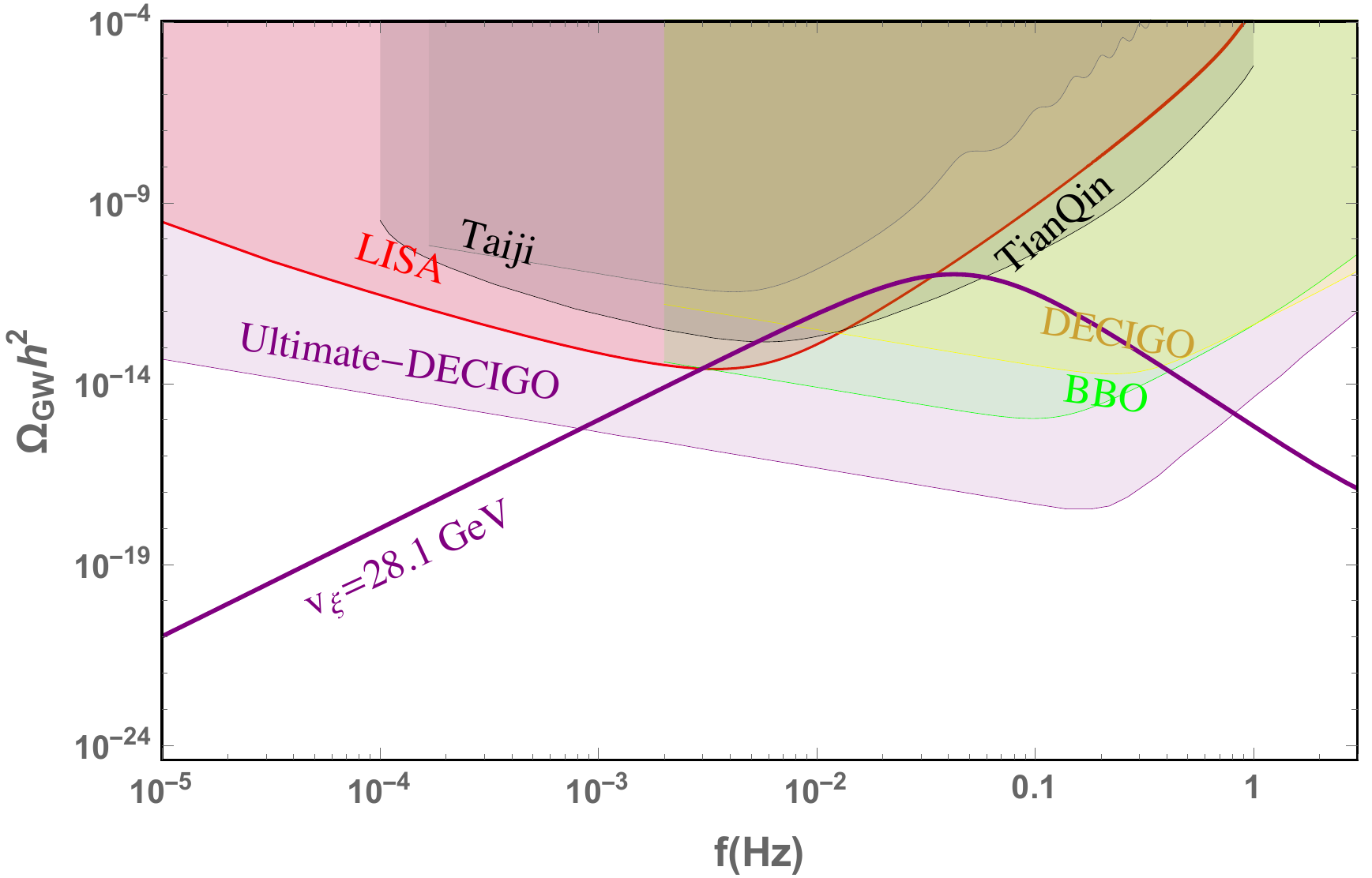}
\includegraphics[width=0.45\textwidth]{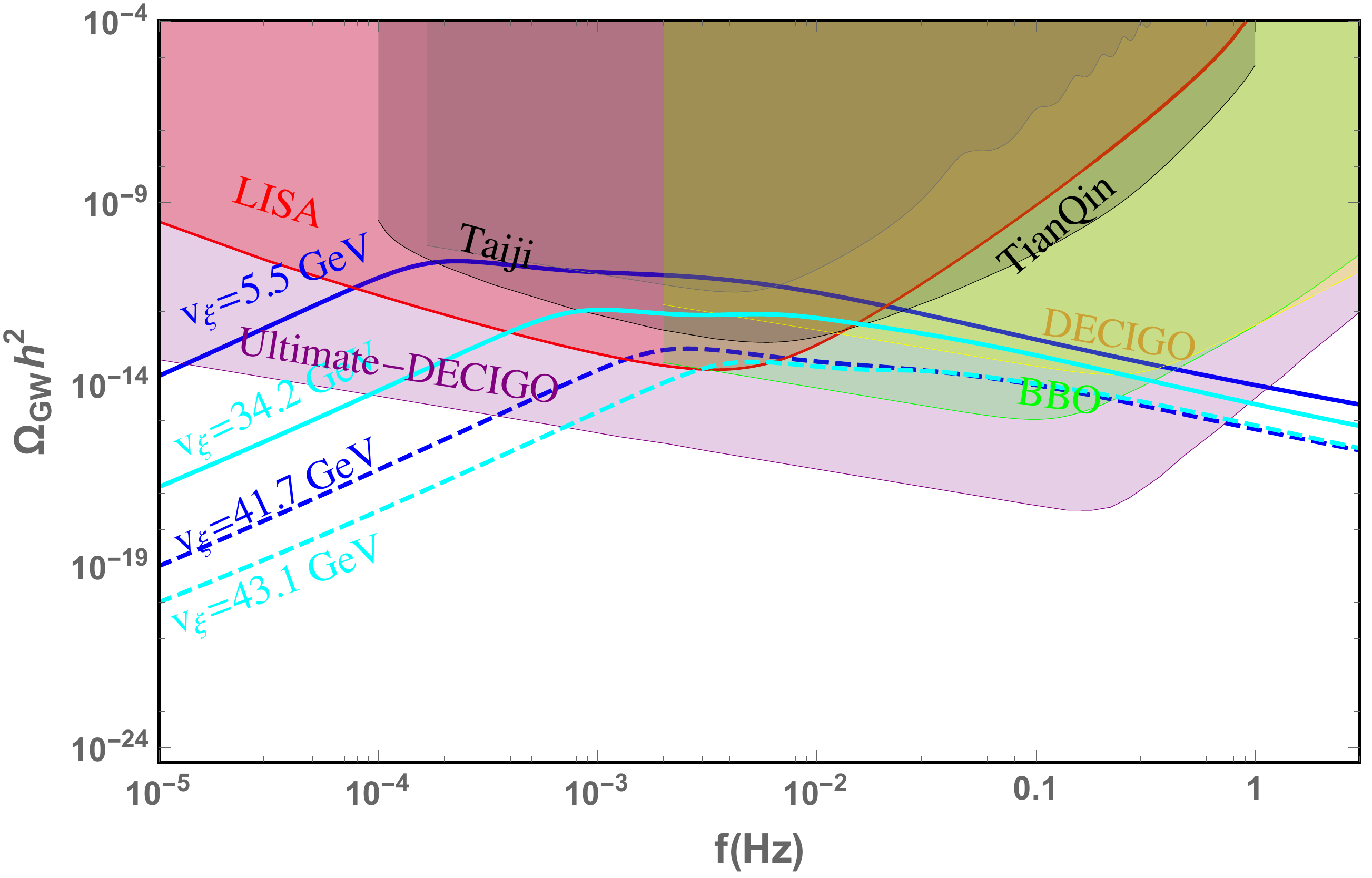}
\end{center}
\caption{
The representative GW signal spectra for \hp{} (left) and \lm{} (right). In \hp{}, only one-step SFOEWPT (purple) is available, 
while in \lm{}, both one-step (cyan) and two-step (blue) are available. The corresponding value of $v_\xi$ for each spectrum is also shown along each line.
}
\label{fig:lowmassGW}
\end{figure}

In~\autoref{fig:lowmassGW}, we show the GW signal spectrum predicted in \hp{} (left) and \lm{} (right) after considering the same-sign diboson bounds from CMS for several representative points taken from the tail of \autoref{fig:betalapha_H5plane} and \autoref{fig:betaalpha_lowmass}(which are the most promising points for GW detection). In comparison with the \hp{}, the GW signals from the SFOEWPT points of \lm{} can be more easy to be probed with a lower frequency. 

With the GW spectrum obtained for each set of parameters input, the GW signals can be searched for using the cross correlation between the outputs of a pair 
of detectors. The detectability of the signals is quantifized by the signal-to-noise ratio(SNR)~\cite{Caprini:2015zlo}:
\begin{eqnarray}
  \text{SNR} = \sqrt{\mathcal{T} \int df 
    \left[
      \frac{h^2 \Omega_{\text{GW}}(f)}{h^2 \Omega_{\text{exp}}(f)} 
  \right]^2} ,
\end{eqnarray}
where $\mathcal{T}$ is the mission duration and $\Omega_{\text{exp}}$ is the power spectral density of a given detector.


\section{Collider Searches}
\label{sec:cols}

\begin{figure}[!tbp]
\centering
\includegraphics[width=\textwidth]{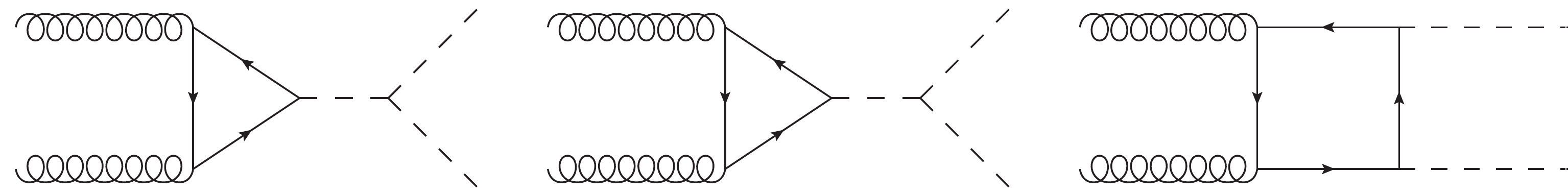}
\put(-460,-5){$g$}
\put(-460,60){$g$}
\put(-300,-5){$g$}
\put(-300,60){$g$}
\put(-140,-5){$g$}
\put(-140,60){$g$}
\put(-330,-5){$h$}
\put(-330,60){$h$}
\put(-170,-5){$h$}
\put(-170,60){$h$}
\put(-10,-3){$h$}
\put(-10,58){$h$}
\put(-365,32){$h$}
\put(-210,32){$H$}
\put(-405,28){$t/b$}
\put(-245,28){$t/b$}
\put(-85,28){$t/b$}
\caption{The Feynman diagrams for the Higgs pair production with extra scalar at the LHC.}
\label{fig:HiggsPairFeynDiag}
\end{figure}

\begin{figure}[!tbp]
\centering
\includegraphics[width=\textwidth]{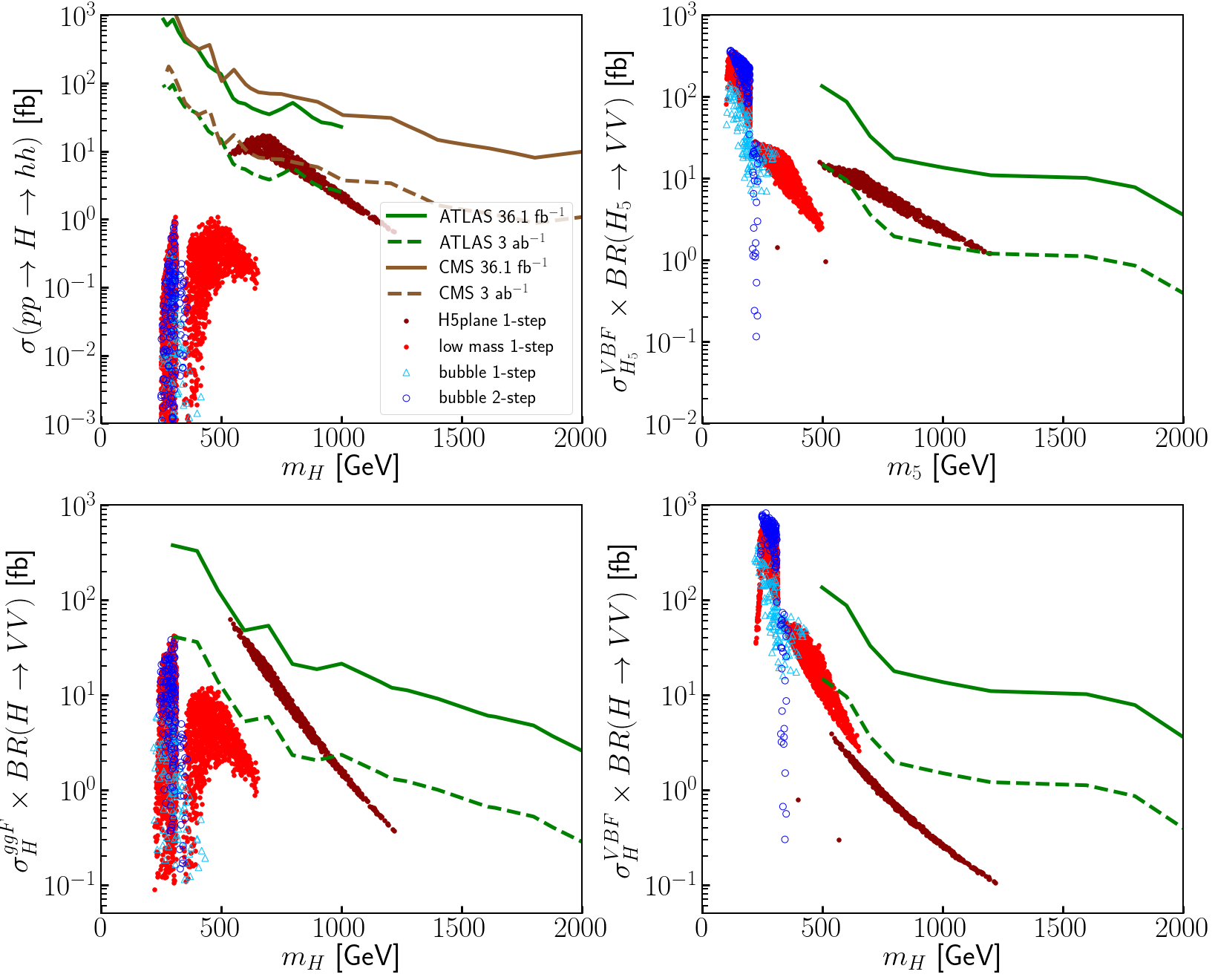}
\caption{The current and prospective di-Higgs and di-boson searches at the LHC for GM model in the H5plane (dark red for one-step) and low mass benchmark (red and blue for one-step and two-step respectively). The blue points are further separated into two categories: ``bubble one-step'' (light blue triangle) and ``bubble two-step'' (dark blue circle).
}
\label{fig:H5plane_lowmass_hh_VV}
\end{figure}

From previous studies (e.g. Fig.~9 of Ref.~\cite{Zhou:2018zli}), we find that one-step and two-step phase transition would happen in different parameter space. 
In the \lm{}, current constraints from same-sign diboson searches already separated the one-step and two-step phase transition into two almost non-overlapping parts with two-step points prefering lower mass while one-step with higher mass. Further, the phase transition strength has a correlation with the triple Higgs coupling in the one-step case, while it is not necessary in the two-step case to have larger triple Higgs coupling to trigger the SFOEWPT. 

On the other hand, the collider searches of the Higgs potential also concentrate on the triple Higgs couplings search: Higgs pairs search at hadron collider (LHC, HL-LHC, SppC, FCC-hh,etc.), $Zhh$ production at lepton colliders (ILC, CEPC, FCC-ee, etc.). Further, the quartic Higgs coupling also enters Higgs pair production through either two-loop~\cite{Bizon:2018syu} or one-loop~\cite{Liu:2018peg} contributions for hadron and lepton collider respectively. Hence the Higgs pair production searches will be nice places to further search these two cases at either the LHC or future lepton colliders.

The leading order contributions for Higgs pair production come from the one-loop diagrams shown in~\autoref{fig:HiggsPairFeynDiag} with non-resonant and possible resonant productions. The non-resonant productions involve the box diagrams (right panel) and also triangle diagram (left panel) which depends on the triple scalar coupling $\lambda_{hhh}$. While the resonant production involve the production of extra scalar and subsequent decays into Higgs pair which depends on the triple scalar coupling $\lambda_{Hhh}$. The differential cross section for the Higgs pair production have been carried out previously~\cite{Eboli:1987dy,Plehn:1996wb}. The calculations have been implemented into {\tt MadGraph} for several different cases~\cite{MG5hh}. For the GM model, we use the NLO UFO model files~\cite{GMFR} implemented with {\tt FeynRules}~\cite{Degrande:2014vpa,Alloul:2013bka} with {\tt MadGraph}~\cite{Alwall:2014hca} to directly calculate the cross section for relevant processes. The required parameter cards are generated using {\tt GMCalc 1.4.1}~\cite{Hartling:2014xma}. Relevant work can also be found in~\cite{Chang:2017niy}.

The Higgs pair production cross section in the GM model are shown in the upper-left panels of Fig.~\ref{fig:H5plane_lowmass_hh_VV} for \hp{} and \lm{}. All points shown in these plots have passed the CMS same-sign diboson searches~\cite{Sirunyan:2017ret,Zhou:2018zli}. For \hp{}, only one-step phase transition exist (dark red points), while both one-step (red points) and two-step (blue points) phase transition can happen for \lm{}. 
For the two-step points (blue points) in \lm{}, we also separate them into two categories: ``bubble one-step'' and ``bubble two-step'', the same as those in~\autoref{fig:betaalpha_lowmass}.
In either case, current Higgs pair searches do not have sufficient sensitivity to probe the phase transition viable parameter space. However, with accumulated data from HL-LHC, it is possible to cover most points in \hp{}. While, points in \lm{} are still beyond the Higgs pair production measurements. 

\begin{figure}[!tbp]
\centering
\includegraphics[width=\textwidth]{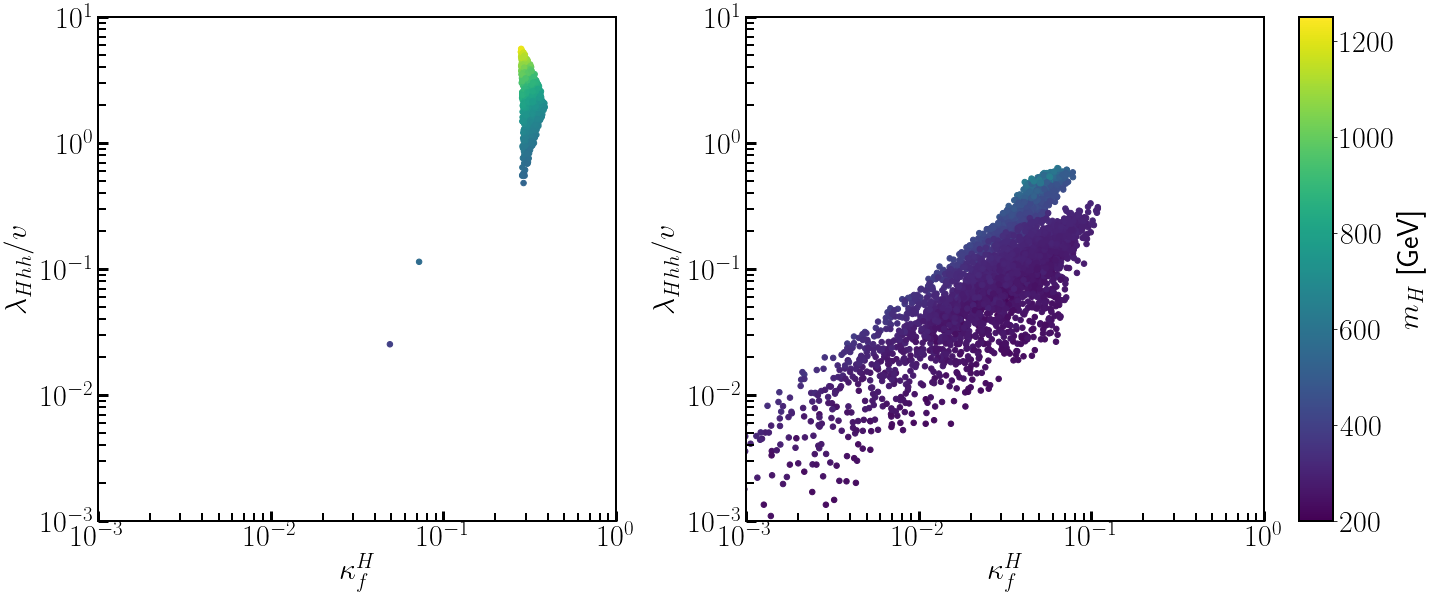}
\caption{SFOEWPT viable points in $\kappa_f^H$-$\lambda_{Hhh}$ plane for H5-plane (left) and low mass benchmark (right). The colors of the points represent the mass of $H$.}
\label{fig:kappaH_LHhh}
\end{figure}

This can be understood from~\autoref{fig:kappaH_LHhh} in which we show the same points in $\lambda_{Hhh}$-BR($H\to hh$) plane for \hp{} (left panel) and \lm{} (right panel). It is clear that in \hp{}, which only has viable one-step points, the triple scalar coupling $\lambda_{Hhh}$ is much larger than those in \lm{}. Hence, the branching of $H\to hh$ will be larger in \hp{}. On the other hand, the $\kappa_f^H\equiv \frac{g_{Hff}^{GM}}{g_{hff}^{SM}} = \frac{\sin\alpha}{\cos\theta_H}$, which contributes to the gluon-gluon fusion cross section, is also larger in \hp{} than that in \lm{}. As a consequence, the cross section of the resonance Higgs pair production is larger in \hp{}.

Beside the Higgs pair measurement, the diboson resonance searches will also have the sensitivity to probe the phase transition viable parameter space. Hence, in both \hp{} as well as \lm{}, we check the resonance diboson cross section through either $H$ (gluon-gluon Fusion and VBF) or $H_5$ (VBF) against the experimental limits. The results are shown in the other three panels in~\autoref{fig:H5plane_lowmass_hh_VV}. From these plots, we find that, \hp{} can be fully covered by the diboson searches from $H_5$ (VBF) and $H$ (ggF) resonance production. While, \lm{} can still escape the searches. However, due to the large $\kappa_V^H$ compared with $\kappa_f^H$, the VBF production of $H$ is not highly suppressed, extending the relevant searches into lower mass region (below 500 GeV) will tremendously improve the sensitivity for such case.

\begin{figure}[!tbp]
\centering
\includegraphics[width=\textwidth]{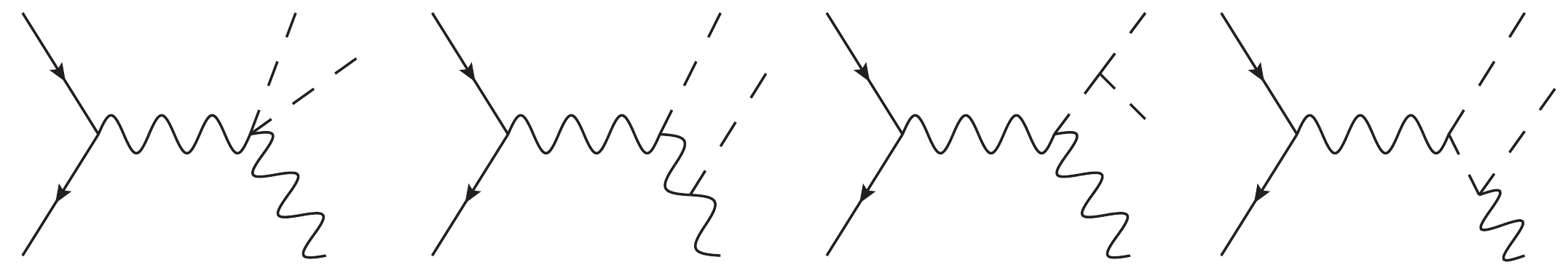}
\put(-470,5){$e^+$}
\put(-470,65){$e^-$}
\put(-420,28){$Z$}
\put(-390,5){$Z$}
\put(-380,65){$h$}
\put(-370,50){$h$}
\put(-350,5){$e^+$}
\put(-350,65){$e^-$}
\put(-300,28){$Z$}
\put(-270,5){$Z$}
\put(-255,60){$h$}
\put(-250,30){$h$}
\put(-230,5){$e^+$}
\put(-230,65){$e^-$}
\put(-180,28){$Z$}
\put(-150,5){$Z$}
\put(-170,50){$h,H$}
\put(-130,65){$h$}
\put(-130,38){$h$}
\put(-115,5){$e^+$}
\put(-115,65){$e^-$}
\put(-65,28){$Z$}
\put(-45,28){$H_3$}
\put(-35,5){$Z$}
\put(-20,60){$h$}
\put(-15,30){$h$}
\caption{The $hhZ$ production at lepton collider in GM model.}
\label{fig:eehhZFeyn}
\end{figure}
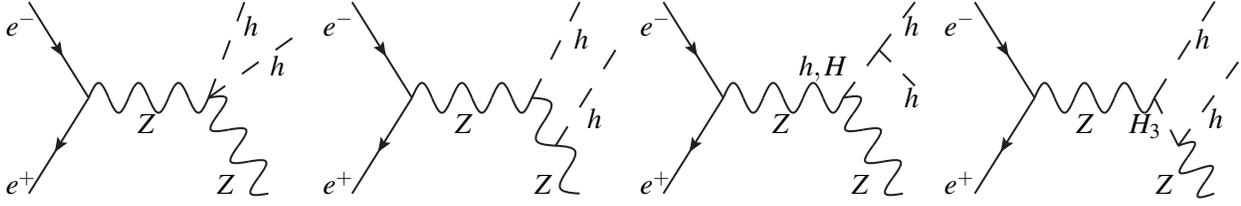

There are also proposals focusing on the electron colliders aiming at the Higgs properties measurements. These lepton colliders also provide another opportunity to search the Higgs pair production~\cite{Li:2017daq}. Hence, we also investigate the sensitivity of the Higgs pair production at the lepton collider associated with Z-boson for \hp{} and \lm{}. The corresponding processes are shown in~\autoref{fig:eehhZFeyn}. The cross sections are calculated using {\tt MadGraph} and the same model files as mentioned above. The SM cross section of such process peaks around $\sqrt{s}=500$ GeV. To maximize the possible sensitivity, we thus focus on the 500 GeV scenario of the lepton collider. 

\begin{figure}[!tbp]
\centering
\includegraphics[width=\textwidth]{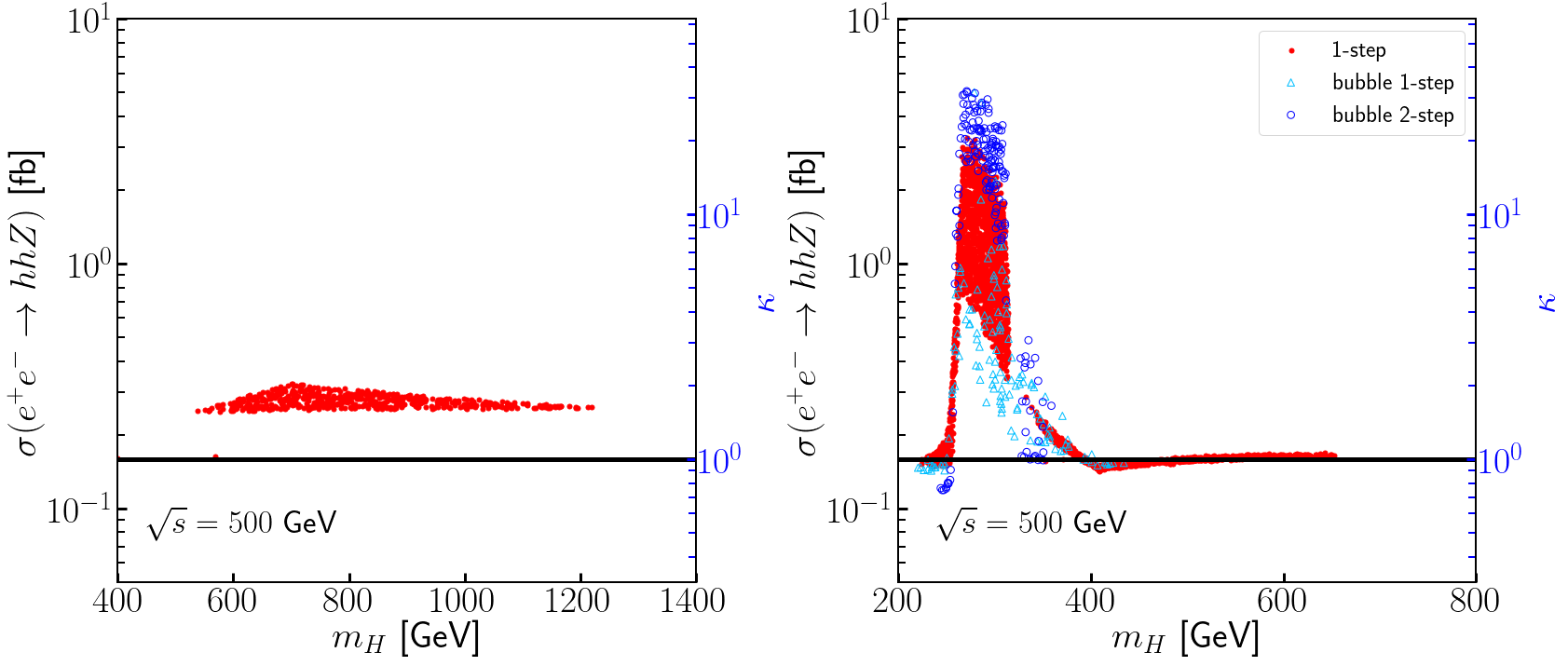}
\caption{The unpolarized cross section of $e^+e^-\to h h Z$ at $\sqrt{s}=500$ GeV for H5-plane (left) and low mass benchmark (right) with one-step (red points) and two-step (blue points) EWPT. The horizontal line indicates the SM value. The secondary y-axis in the right-hand side of each plot indicates $\kappa\equiv\frac{\sigma^{GM}}{\sigma^{SM}}$. The blue points are further separated into two categories: ``bubble one-step'' (light blue triangle) and ``bubble two-step'' (dark blue circle).
}
\label{fig:eehhZILC500}
\end{figure}

The unpolarized total cross sections are shown in~\autoref{fig:eehhZILC500} for \hp{} (left panel) and \lm{} (right panel). The enhancement factor $\kappa = \frac{\sigma^{GM}}{\sigma^{SM}}$ is also indicated in the secondary y-axis in the right-hand side of each plot. We find that in \hp{}, the cross section has a moderate enhancement with $\kappa\sim 2$ for the entire viable points. However, in \lm{}, $\kappa$ spans a large range, and can even reach about 30 for mass around 300 GeV. 
For \hp{}, the moderate enhancement mainly comes from the large $\lambda_{hhh}$, as the collision energy is not enough for the resonance production through either $H$ or $H_3$. While, in the low mass benchmark, these masses are within the reach of the collision energy, the resonance production of $H$ and/or $H_3$ induce the huge enhancement.

\begin{figure}[!tbp]
\centering
\includegraphics[width=\textwidth]{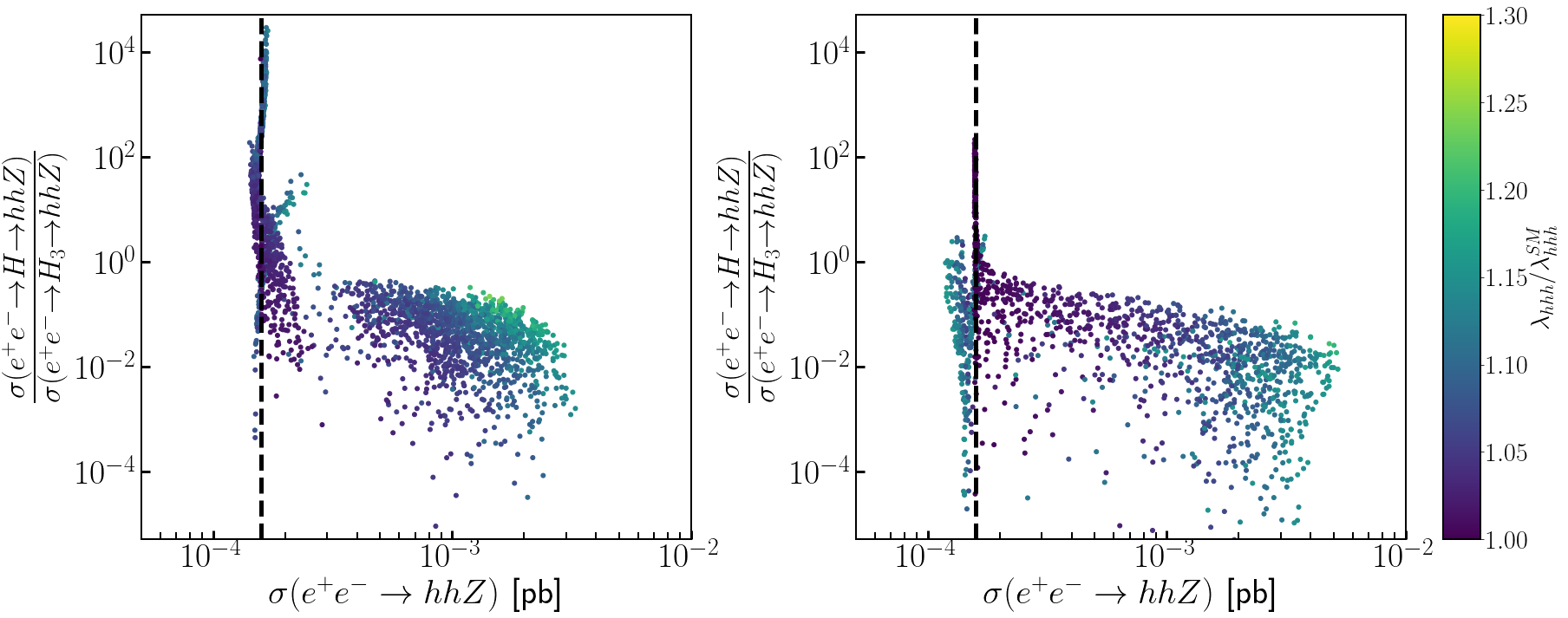}
\caption{The ratio between $H$ and $H_3$ induced the cross section with respect to the total cross section in \lm{} for one-step (left) and two-step (right) SFOEWPT.}
\label{fig:eehhZ_H_H3}
\end{figure}

\autoref{fig:eehhZ_H_H3} shows the ratio between $H$ and $H_3$ induced the cross section with respect to the total cross section in \lm{} for both one-step (left) and two-step (right) cases. It is clear that the enhancement in the total cross section mainly comes from $H_3$ resonance in \lm{}. When the collision energy raise to even higher value, it is also possible, in \hp{}, to enhance the Higgs pair production through $H_3$ and/or $H$ resonance.
With the high precision that we can achieve for the cross section measurement at the ILC/CEPC, these machines will have sensitivity for these SFOEWPT viable points.

\begin{figure}[!tbp]
\centering
\includegraphics[width=\textwidth]{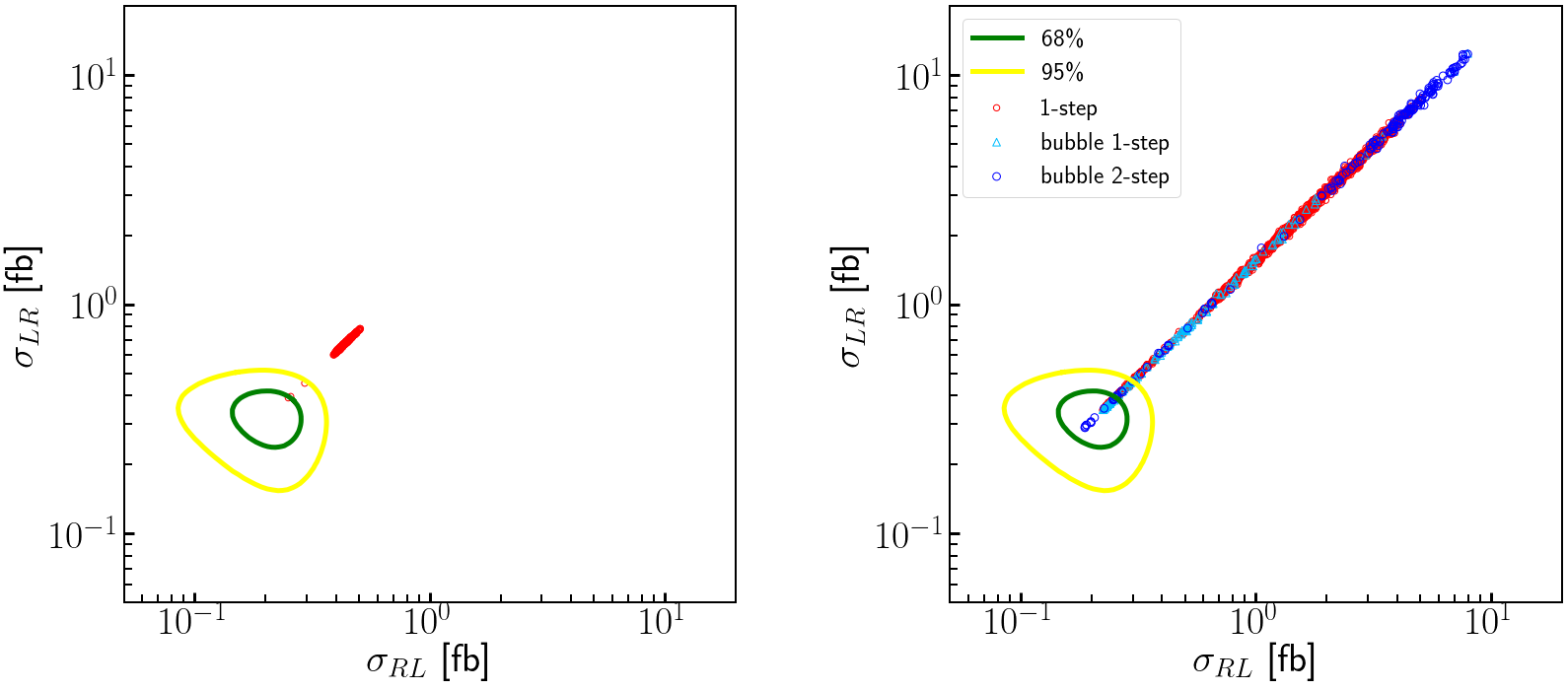}
\caption{The constraints of the polarized $hhZ$ production cross section measurement in the $\sigma_{RL}$-$\sigma_{LR}$ plane for H5-plane (left) and low mass benchmark (right). The red and blue points are the one-step and two-step viable points in the GM parameter space respectively. The green and yellow lines are the 68\% and 95\% C.L. contour respectively. The blue points are further separated into two categories: ``bubble one-step'' (light blue triangle) and ``bubble two-step'' (dark blue circle).
}
\label{fig:eehhZPolarization}
\end{figure}

To give a more concrete sensitive study of such channel, the polarized cross sections are also calculated for $P(e^-,e^+)=(-80\%,30\%)$ and $P(e^-,e^+)=(80\%,-30\%)$. We combine the sensitivities from $bbbb$ and $bbWW$ channels of this process from~\cite{Duerig:2016dvi,LC-REP-2013-025} to obtain the constraints in the $\sigma_{RL}$-$\sigma_{LR}$ plane which is shown in~\autoref{fig:eehhZPolarization}. From this plot, we find that the almost all the SFOEWPT viable points in \hp{} can be excluded by this measurement. In the low mass benchmark, most points can also be excluded. However, we still have both one-step and two-step SFOEWPT viable points that are beyond the sensitivity in \lm{}.

Some other channels are also possible as a complimentary to the GW signal, especially for lower mass region. In~\cite{Logan:2018wtm}, authors studied the sensitivities from the loop-induced channel $W\gamma$ from fermiophobic scalar which is specific for \lm{}. On the other hand, the diphoton searches~\cite{Aad:2014ioa,Aaboud:2017yyg} are also promising. However, after re-interpret the bounds from~\cite{Logan:2018wtm} in our case, we found that they are not yet sensitive to reach the SFOEWPT viable points. Further improvement and detailed studies are needed for these searches.

\begin{figure}[!tbp]
\centering
\includegraphics[width=\textwidth]{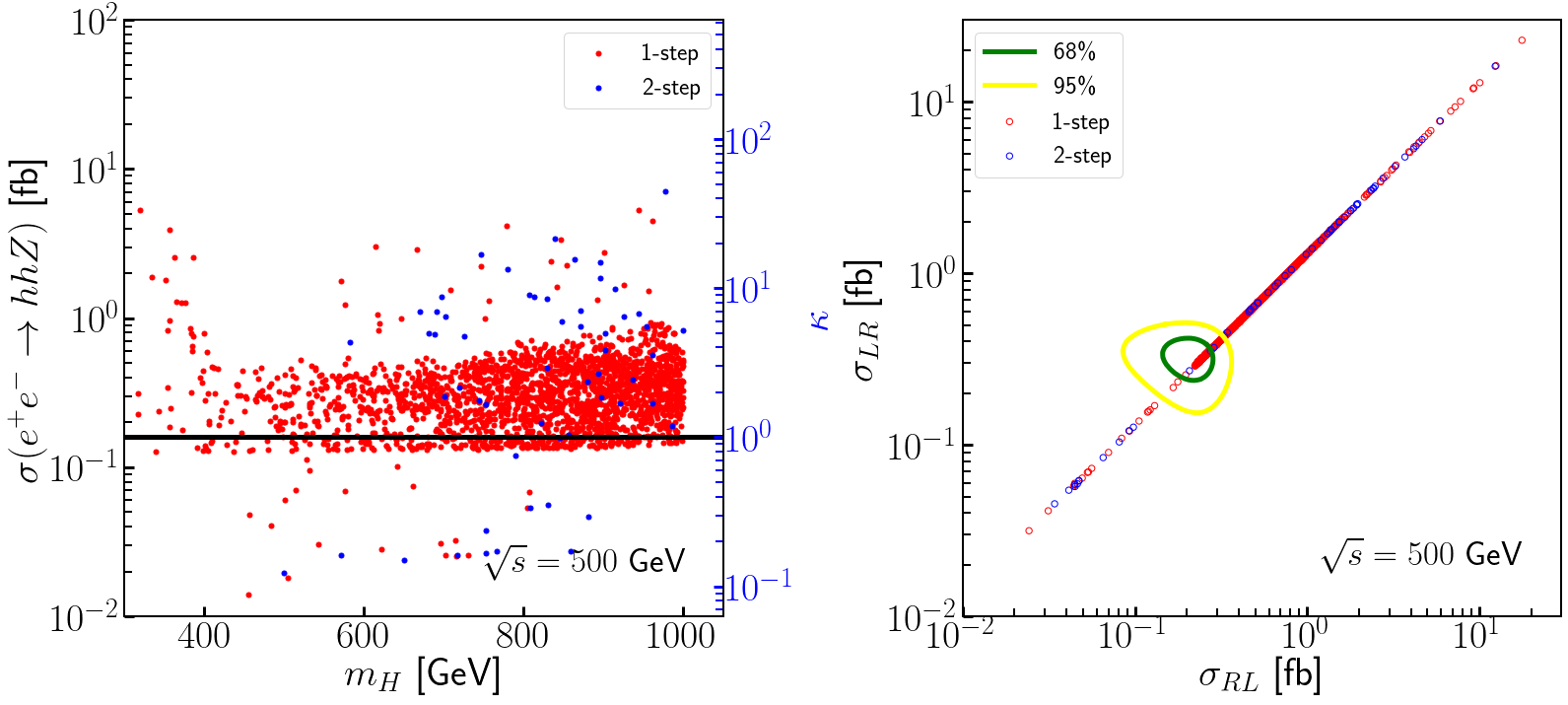}
\caption{The unpolarized cross section of $e^+e^-\to h h Z$ at $\sqrt{s}=500$ GeV (left) and the constraints of the polarized $hhZ$ production cross section measurement in the $\sigma_{RL}$-$\sigma_{LR}$ plane (right) for xSM.}
\label{fig:eehhZ_xSM}
\end{figure}

In xSM, the high mass region of the extra Higgs is not going to be covered by Higgs pair production searches at the future HL-LHC, and a lot parameter spaces there are not going to be probed by the diboson searches due to the small mixing angle of the SM Higgs and the heavy extra Higgs suppress effect. These regions can be complementary searched by the gravitational wave space-based detectors~\cite{Alves:2018jsw}. At the ILC, the Higgs pair search results are shown in~\autoref{fig:eehhZ_xSM}, where the cross sections are obtained using the same method as in GM model. We find that in xSM model, the cross sections (unpolarized or polarized) has moderate enhancement. From the prospects of the future cross section measurement, we could exclude most one-step as well as two-step points in xSM.

\begin{figure}[!tbp]
\begin{center}
\includegraphics[width=\textwidth]{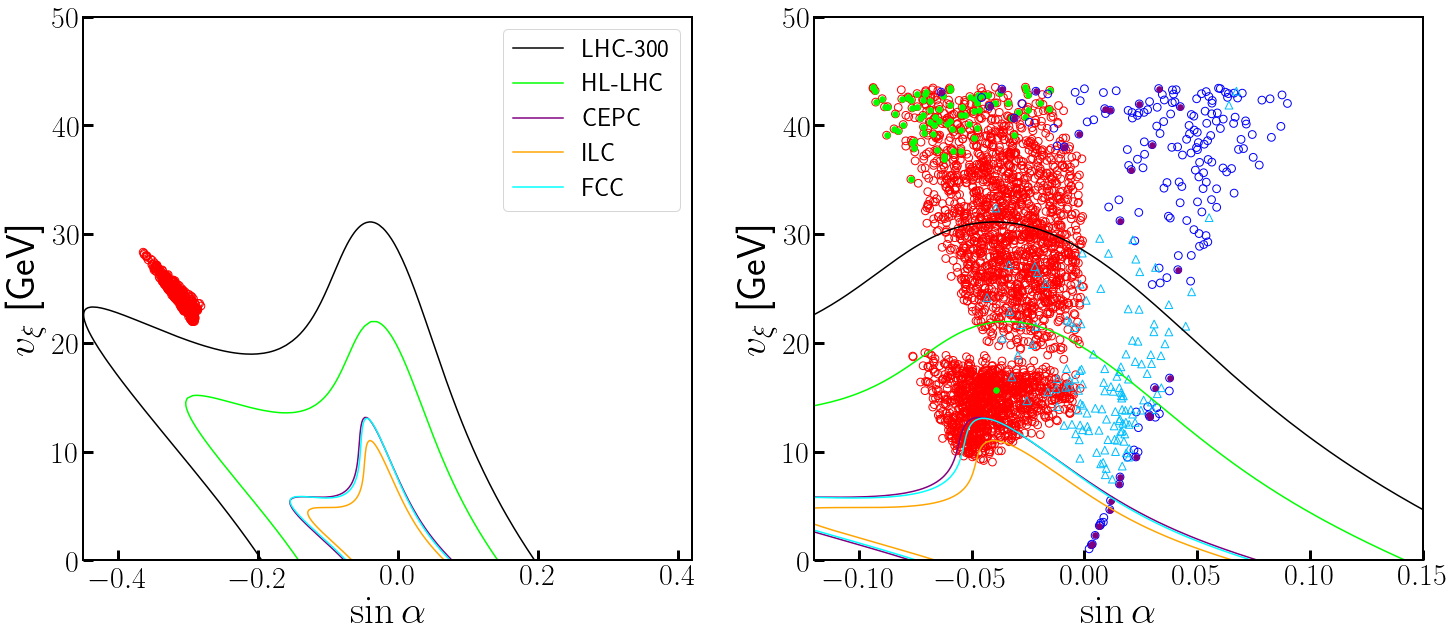}
\end{center}
\caption{
The SFOEWPT viable points (bubble nucleation can occur and $v_n/T_n>1$) in the $\sin\alpha$-$v_{\xi}$ plane for one-step (red) and two-step (blue) phase transition for \hp{} (left panel) and \lm{} (right panel) scenario. The contours with different colors represent the constraints from the Higgs precision measurement from different experiments as indicated in the legend. In \lm{}, we also indicate the points having $SNR>10$ with solid markers for one-step (green) and two-step (purple) cases.
}
\label{fig:FurtureConstraintsGM}
\end{figure}

In~\autoref{fig:FurtureConstraintsGM}, we show the future hadron and lepton colliders sensitivity to the parameter spaces of one-step and two-step SFOEWPT for \hp{} and \lm{} from the Higgs signal strength measurement, the conventions are the same as those in~\autoref{fig:betaalpha_lowmass}. 
Here, all the curves are obtained by fitting the SM-like Higgs signal strength measurement from each experiments prospects (LHC~\cite{ATL-PHYS-PUB-2014-016,ATL-PHYS-PUB-2013-014,ATL-PHYS-PUB-2014-012,ATL-PHYS-PUB-2014-006,ATL-PHYS-PUB-2014-011,ATL-PHYS-PUB-2014-018,Hartmann:2015aia,ATLAS-collaboration:1484890}, CEPC~\cite{CEPCStudyGroup:2018ghi}, ILC~\cite{Bambade:2019fyw} and FCC-ee~\cite{Mangano:2018mur,Benedikt:2018qee}).
From~\autoref{fig:FurtureConstraintsGM}, we find that, most one-step points in both \hp{} and \lm{} can be excluded by the Higgs signal strength measurements. While we still have a bunch of points for the two-step case locating around the alignment limit and escaping the signal strength measurements. 
The stochastic GWs can be detected by finding the cross correlation of two independent interferometers (SNR)~\cite{Caprini:2015zlo}, we calculate the quantity for LISA and the $SNR>10$ points mostly concentrate in the region of a large $v_\xi$ of the  \lm{}. 

\begin{figure}[!tbp]
\begin{center}
\includegraphics[width=0.9\textwidth]{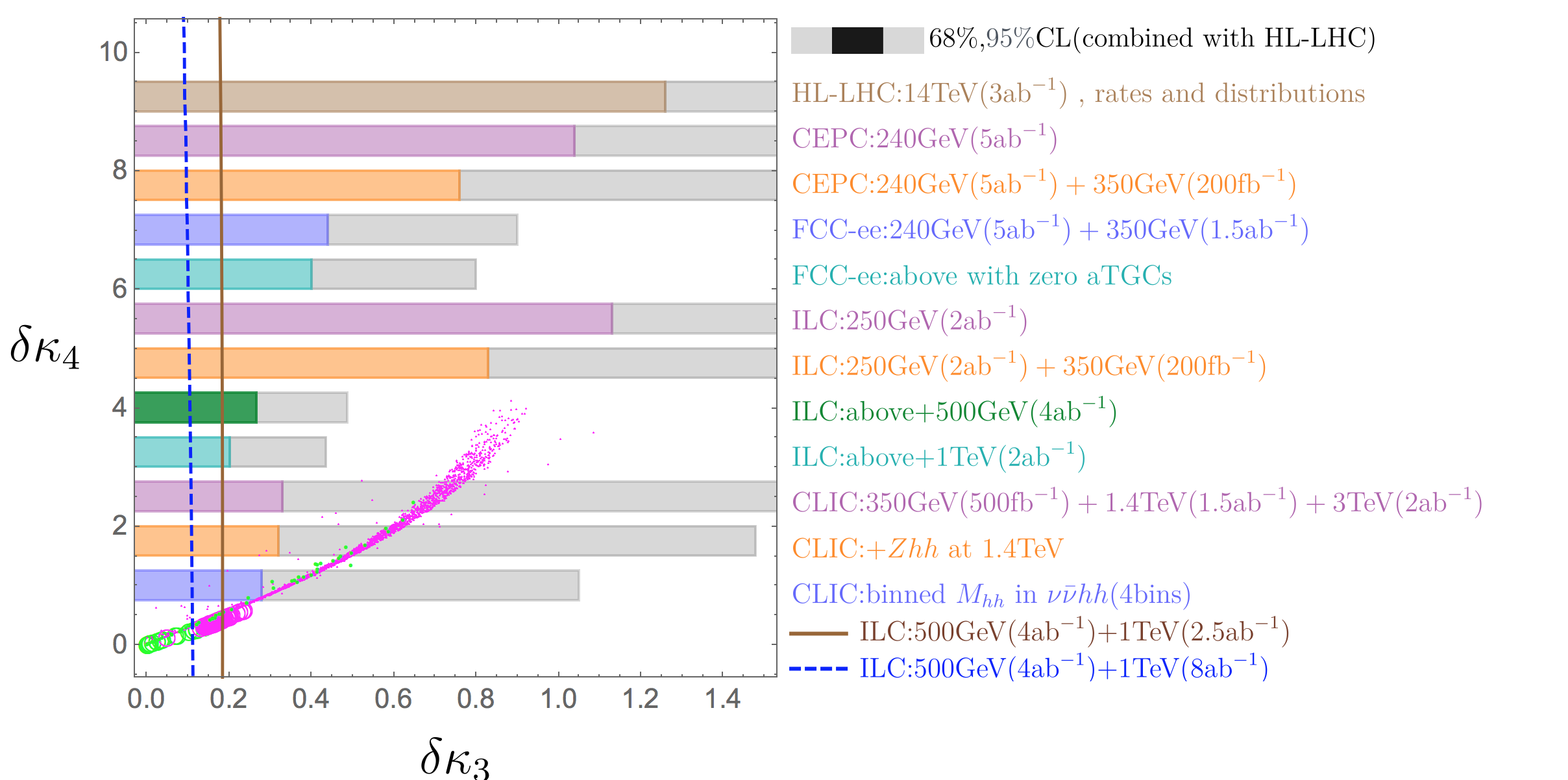}
\caption{Collider sensitivities of the triple Higgs coupling and quartic higgs coupling with the $SNR>10$ points. 
  The green and magenta represent the two-step and one-step SFOEWPT points. 
  The circle points and the dotted points represent the GM and xSM model scenarios.
  The bars and contours are the sensitivities taken from Ref.~\cite{DiVita:2017vrr,Liu:2018peg}.
}
\label{fig:k3k4}
\end{center}
\end{figure}

We finally present the future colliders sensitivities in the SFOEWPT parameter spaces, with an conservative consideration where only the deviation of the SM couplings are included.
We first show in~\autoref{fig:k3k4} the measurements of the two couplings ($\lambda_{hhh}$ and $\lambda_{hhhh}$) at future $e^+e^-$ colliders and the HL-LHC. Ref.~\cite{DiVita:2017vrr} provides the $68\%$CL and $95\%$CL results (the inner and outer horizontal bar regions) of the future $e^+e^-$ colliders and HL-LHC measurements, though which only include the cubic coupling in their analysis based on effective field theory approach. 
Another future ILC measurements precision (brown and blue lines) are adopted from Ref.~\cite{Liu:2018peg}, where the analysis are also based on effective field theory approach, and therefore we expect the UV model search would provide much batter prediction. 
For the situation with additional EWSB contributions, i.e., the GM model, we expect the future lepton hadron collider be more powerful, since the phase transition possibility would be highly restricted by collider searches, especially the diboson search. 

\section{Conclusions}
\label{sec:conc}

Utilizing the Georgi-Machacek (GM) model, we simultaneously study Higgs phenomenology and the Electroweak phase transition for the scenario with the eEWSB contribution. The eEWSB is found to be helpful to realize different vacuum structure from the SM as the Universe cools down, and therefore leads to different strongly first order Electroweak phase transition patterns. We explore two benchmarks of the Georgi-Machacek (GM) to demonstrate the point.  For both the one-step and the two-step SFOEWPT, we performed the study of the gravitational wave searches and the Higgs pair searches of the cubic Higgs couplings at hadronic and leptonic colliders. For comparison, we present the xSM model to show the situation without eEWSB. We found that:  1) most of the SFOEWPT parameter spaces is able to be covered by the future lepton colliders;  2) the GWs mostly come from the second-step SFOEWPT in the GM model. 
3) in comparison with the xSM model, the GW signals (to be probed by LISA ) from the SFOEWPT of the GM model requires much higher precision of the colliders.

\section*{Acknowledgment}

We thank Yi Liao, Zhi-Long Han and Bin Li for helpful discussions. The work of LGB is Supported by the National Natural Science Foundation of China (under grant No.11605016 and No.11647307), Basic Science Research
Program through the National Research Foundation of Korea (NRF) funded by the Ministry of
Education, Science and Technology (NRF-2016R1A2B4008759), and Korea Research Fellowship Program through the National Research Foundation of Korea (NRF) funded by the Ministry
of Science and ICT (2017H1D3A1A01014046). The work of Y.C.W. is partially supported by the
Natural Sciences and Engineering Research Council of Canada. H.G. is supported in part by the
U.S. Department of Energy grant ds-sc0009956.

\newpage

\appendix

\section{The GM model}
\label{app:GMmodel}
In the Georgi-Machacek model, there are one isospin doublet scalar field $\phi = (\phi^+,\phi^0)^T$ with hypercharge $\rm Y=\frac{1}{2}$, one complex isospin triplet scalar field $\chi = (\chi^{++},\chi^+,\chi^0)^T$ with hypercharge $\rm Y=1$, and one real triplet $\xi = (\xi^+,\xi^0,-\xi^{+*})^T$ with hypercharge $\rm Y=0$. The custodial symmetry is introduced at tree level by imposing a global SU(2)$_L\times$SU(2)$_R$ symmetry upon the scalar potential. 
The neutral components of each fields can be parameterized into real and imaginary parts according to
\begin{eqnarray}
\rm \phi^0 = \frac{\nu_{\phi} + h_\phi + i a_\phi}{\sqrt{2}},
\rm \qquad
\rm \chi^0 = \frac{\nu_{\chi} + h_\chi + i a_\chi}{\sqrt{2}},
\rm \qquad
\rm \xi^0 = \nu_{\xi} + h_\xi,
\end{eqnarray}
where $\nu_\phi$, $\nu_\chi$ and $\nu_\xi$ are the VEVs of $\phi^0$, $\chi^0$ and $\xi^0$, respectively.
With only neutral components of this model, the potential reads:
\begin{align}
V_{0}&= \frac{1}{4}(4 h_{\phi}^{4} \lambda_1 + 2 (h_{\xi}^2+h_{\chi}^2)(m_2^2+2\lambda_2 (h_{\xi}^2+h_{\chi}^2))+2\lambda_3(2 h_{\xi}^4 + h_{\chi}^4)\nonumber\\
&+ h_{\phi}^2 (2 m_1^2 + 4  \lambda_4 h_{\xi}^2 + h_{\xi}(2\sqrt{2} \lambda_5 h_{\chi}  + \mu_1) + h_{\chi}(4 \lambda_4   h_{\chi}+\lambda_5  h_{\chi} + \sqrt{2} \mu_1)) + 12 \mu_2  h_{\xi} h_{\chi}^2 ).
\label{eq:Vtree_appendix}
\end{align}
We can derive the EWSB vacuum through the minimization conditions:
\begin{eqnarray}
\frac{\partial V_{0}}{\partial h_{\phi}} \ = \ \frac{\partial V_{0}}{\partial h_{\chi}}
\ = \ \frac{\partial V_{0}}{\partial h_{\xi}} \ = \ 0 ~,
\label{equ:vacuum}
\end{eqnarray}
where the fields other than
$\phi^0$, $\chi^0$ and $\xi^0$
take zero VEV's. In this paper, the solution satisfying the relation $v_{\chi}=\sqrt{2}v_{\xi}$ is selected, by which the EWSB vacuum maintains the diagonal $SU(2)_V$ symmetry. Thus the parameter $\rho = m_W^2/(m_Z^2 \cos\theta_w^2) = 1$ is established at the tree level.
The W and Z boson masses from the EWSB give the constraint,
\begin{equation}
\nu_\phi^2 + 8\nu_\xi^2 \equiv \nu^2 = \frac{1}{\sqrt{2}G_F}\approx (246\text{ GeV})^2\;.
\end{equation}

When $\nu_\phi,\nu_\xi \neq 0$, with the help of~\autoref{equ:vacuum} (under the relation $\nu_\chi = \sqrt{2}\nu_\xi$), we could rewrite $m_1^2$, $m_2^2$ in terms of $\nu_\phi$,$\nu_\xi$ and other parameters in the Higgs potential as:
\begin{align}
m_1^2 \ &= \ -4\lambda_1 \nu_{\phi}^2 - 6\lambda_4 \nu_\xi^2 - 3\lambda_5 \nu_\xi^2
- \frac{3}{2} \mu_1 \nu_\xi \;,\\
m_2^2 \ &= \ -12\lambda_2 \nu_\xi^2 - 4\lambda_3 \nu_\xi^2 - 2\lambda_4 \nu_{\phi}^2
- \lambda_5 \nu_{\phi}^2 - \mu_1 \frac{\nu_{\phi}^2}{4\nu_\xi} - 6\mu_2 \nu_\xi\;.
\label{m1m2}
\end{align}
There are 13 scalar fields in this model.
After diagonalizing the mass matrices, the fields can be rewritten as the physical scalars (quintuple, triplet and singlet respectively)
\begin{align}
& H_5^{++} = \chi^{++} ~,\quad H_5^+ = \frac{1}{\sqrt{2}} \bigg( \chi^+ - \xi^+ \bigg) ~,\quad  H_5^0 = \sqrt{\frac{1}{3}} h_{\chi} - \sqrt{\frac{2}{3}} h_{\xi} ~,\\
&H_3^+ = -\cos \theta_H \, \phi^+ + \sin \theta_H \, \frac{1}{\sqrt{2}} \bigg( \chi^+ + \xi^+ \bigg) ~,
\quad  H_3^0 = -\cos \theta_H \, a_{\phi} + \sin \theta_H \, a_{\chi} ~, \\
&h = \cos \alpha \, h_{\phi} - \frac{\sin \alpha}{\sqrt3} \, \bigg( \sqrt{2} h_{\chi} + h_{\xi} \bigg) ~,\quad H_1 = \sin \alpha \, h_{\phi} + \frac{\cos \alpha}{\sqrt3} \, \bigg( \sqrt{2} h_{\chi} + h_{\xi} \bigg) ~,
\end{align}
and the goldstone bosons
\begin{align}
& G^+ = \sin\theta_H \phi^+ + \cos\theta_H \frac{1}{\sqrt2}(\chi^+ + \xi^+)~, ~~G^0 = \sin\theta_H a_\phi + \cos\theta_H a_\xi\;,
\end{align}
where $\sin\theta_H = \frac{2\sqrt{2}\nu_\xi}{\nu}$ and $\cos\theta_H = \frac{\nu_\phi}{\nu}$, and $\alpha$ is the mixing angle between two singlets which is determined by the mass matrix of these scalars as will be shown below.

The 3 goldstone bosons eventually become the longitudinal components of the W and Z bosons, while, the remaining 10 physical fields can be organized into a quintuple $H_5$ $=$ $(H^{++}_5$, $H^+_5$, $H^0_5$, $H^-_5$, $H^{--}_5)^T$, a triplet $H_3 = (H^+_3, H^0_3, H^-_3)^T$ and two singlets $h$ and $H_1$, where the former ($h$) is used to denote the SM-like Higgs boson. The triplet scalar is CP-odd, while others are CP-even.
The masses of different multiplets can be written as
\begin{align}
m^2_{H_5}=&m^2_{H_5^{\pm\pm}}=m^2_{H_5^{\pm}}=m^2_{H_5^0} = (8 \lambda_3 \nu^2_{\xi}- \frac{3}{2}\lambda_5 \nu^2_{\phi})  - \frac{\mu_1 \nu^2_{\phi}}{4 \nu_\xi} - 12 \mu_2\nu_\xi \;,\\
m^2_{H_3}=&m^2_{H_3^\pm}=m^2_{H_3^0} = - (\frac{\lambda_5}{2} + \frac{\mu_1}{4\nu_\xi})\nu^2\;.
\end{align}
The singlets masses of $m_{h,H_1}$ are the eigenvalues of mass matrix written in terms of gauge eigenstates:
\begin{align}
M^2 = \left(\begin{array}{cc}
M^2_{11} & M^2_{12} \\
M^2_{12} & M^2_{22}
\end{array}\right)\;,
\end{align}
with
\begin{align}
M^2_{11} &= 8 \cos^2\theta_H \lambda_1 \nu^2,\\
M^2_{22} &= \sin^2\theta_H (3\lambda_2+\lambda_3)\nu^2 + \cos^2\theta_H M^2_1-\frac{1}{2}M_2^2, \\
M^2_{12} &= \sqrt{\frac{3}{2}}\sin\theta_H \cos\theta_H[(2\lambda_4+\lambda_5)\nu^2-M_1^2],
\end{align}
where $M_1^2 = -\frac{\nu}{\sqrt{2}\sin\theta_H}\mu_1$ and $M_2^2=-3\sqrt{2}\sin\theta_H \nu \mu_2$. The mixing angle $\alpha$ is determined by
\begin{equation}
\tan2\alpha = \frac{2M^2_{12}}{M^2_{22}-M^2_{11}}\;,
\end{equation}
as a function of the $\theta_H$.

\section{EWPT in the SM + Real Singlet: xSM model}

For the xSM model, the gauge invariant finite temperature effective potential is found to be~\cite{Profumo:2014opa,Kozaczuk:2015owa,Alves:2018jsw}:
\begin{eqnarray}
 && V(h,s,T) = - \frac{1}{2} [\mu^2 - \Pi_h(T)] h^2
  - \frac{1}{2} [-b_2 - \Pi_s(T)] s^2 \nonumber \\
  &&\hspace{0.7cm} + \frac{1}{4} \lambda h^4 + \frac{1}{4} a_1 h^2 s + \frac{1}{4} a_2 h^2 s^2 +
  \frac{b_3}{3} s^3 + \frac{b_4}{4} s^4, \quad \quad
\end{eqnarray}
with the thermal masses given by
\begin{eqnarray}
  &&  \Pi_h(T) =\left( \frac{2 m_W^2 + m_Z^2 + 2 m_t^2}{4 v^2} + \frac{\lambda}{2} + \frac{a_2}{24} \right) T^2, \nonumber \\
  &&  \Pi_s(T) =\left( \frac{a_2}{6} + \frac{b_4}{4} \right) T^2,
\end{eqnarray}

Identifying the coefficient for terms with the same power of different fields, we could get the correspondence of the parameters in GM and xSM as listed in~\autoref{tab:conv}.

\begin{table}[!tbp]
\caption{xSM convention vs GM convention}
\label{tab:conv}
\begin{center}
\begin{tabular}{c c c }
\hline
~~~~~~~~~~~xSM~~~~~~~~~~& ~~~~~~~~~GM~~~~~~~~~ \\
\hline
  $\frac{1}{24}T^2(a_2+3(3 g^2+g'^2+2 Y_t^2+4\lambda))-\mu^2$ ~~&~~ $\frac{T^2}{8}(3 g^2+g'^2+16\lambda_1+12\lambda_4+Y_t^2 \sec\theta_H)+m_1^2$  \\
  $b_2+\frac{1}{12}(2a_2+3b_4)T^2$ ~~&~~ $3m_2^2+T^2(3 g^2+g'^2+11\lambda_2+7\lambda_3+2\lambda_4)$  \\
  $\lambda$ & $4\lambda_1$ \\
  $b_4$ & $12(3\lambda_2 + \lambda_3)$  \\
  $a_2$  &$6(2\lambda_4+\lambda_5)$ \\
  $a_1$ & $3 \mu_1$  \\
  $b_3$ & $18 \mu_2$  \\
\hline
\end{tabular}
\end{center}
\end{table}

\section{The \texorpdfstring{$h^6$}{h6} operator for GM and xSM models}

The dimensional six operator $h^6$ can modify both cubic and quartic Higgs couplings, with the Higgs potential for the Higgs boson given by:
\begin{equation}
V_{SM+h^6}^h=-\frac{\mu_h^2}{2} h^2 +\frac{\lambda_h}{4} h^4+\frac{1}{8\Lambda^2} h^6\,.
\label{Vdim6}
\end{equation}
Here, the $\Lambda$ indicates the scale where the heavy particles are integrated out.
The minimization conditions lead to the following relations,
\begin{equation}
\mu_h^2=\frac{m_h^2}{2}-\frac{3 v^4}{4 \Lambda^2}\;,
\lambda=\frac{m_h^2}{2 v^2}-\frac{3 v^2}{2\Lambda^2}\;.
\end{equation}
In this scenario, the requirement that the EW minimum being the global minimum leads to 
\begin{equation}\label{eq:Lam}
\Lambda\geq v^2/m_h\;.
\end{equation}
The cubic and quartic Higgs couplings are modified by the dimension-6 operator as
\begin{equation}
\lambda^{h^6}_{3h}=\frac{3 m_h^2}{v}+\frac{6 v^3}{\Lambda^2}\;,
\lambda^{h^6}_{4h}=\frac{3 m_h^2}{v^2}+\frac{36 v^2}{\Lambda^2}\;.
\end{equation}
The phase transition dynamics are estimated after taking into account the thermal correction of $c_h$, as in Ref.~\cite{Huang:2015tdv}.
As studied in Ref.~\cite{Grojean:2004xa,Huang:2015tdv}, to obtain a SFOEWPT with the $h^{6}$ operators additional to the SM, the additional contribution to the quartic Higgs coupling $\lambda$ should be compensate it's negative value and therefor to ensure the possibility of the SFOEWPT. 

\begin{figure}[!htp]
\begin{center}
\includegraphics[width=0.6\textwidth]{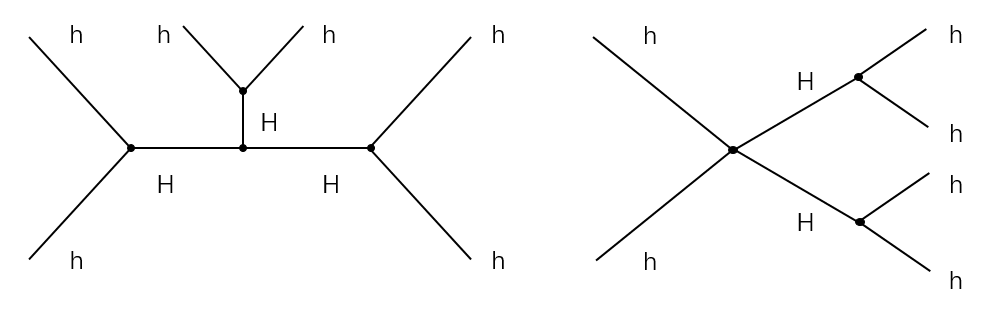}
\end{center}
\caption{The Feynnman diagram for $(H^\dag H)^3$ operator.}\label{fig:h6dia}
\end{figure}

Suppose the extra Higgs is much heavy than the SM Higgs, the $(H^\dag H)^3$ operator can be obtained after integrated out heavy Higgs in~\autoref{fig:h6dia}.
For the GM model, the relevant interactions are given by,
\begin{eqnarray}
\lambda^{GM}_{2hH}&&=\frac{\sqrt{3}}{2}  \cos \alpha (\mu_1  \cos \alpha ^2-2 \sin \alpha  ^2 (\mu_1-4 \mu_2)) \nonumber \\
\lambda^{GM}_{2h2H}&&=\cos 4 \alpha (-3 \lambda_1 -3 \lambda_2-\lambda_3+3 \lambda_4+\frac{3 \lambda_5}{2})+3 \lambda_1+3 \lambda_2+\lambda_3+\lambda_4+\frac{\lambda_5}{2}\nonumber \\
\lambda^{GM}_{3H}&&=\frac{ \sqrt{3}}{2} \cos \alpha (3 \mu_1 \sin \alpha ^2+8 \mu_2 \cos \alpha ^2) .
\end{eqnarray}
Using the equation of motion on~\autoref{eq:GMalp}, the coefficients of $c_4$,$c_6$, $c_8$ for the GM are obtained as follows,
\begin{eqnarray}
c^{GM}_4&=&\lambda_1  \cos^4 \alpha+\sin^4 \alpha (\lambda_2+\frac{\lambda_3}{3})+\lambda_4 \sin^2\alpha \cos^2\alpha+\frac{1}{8} \lambda_5 \sin^2 2 \alpha  \nonumber\\
&&-\frac{3  \cos ^2\alpha (\cos 2 \alpha  (8 \mu_2-3 \mu_1)+\mu_1-8 \mu_2)^2}{128 (m_1^2 \sin ^2\alpha+m_2^2 \cos^2 \alpha )}\nonumber\\
c^{GM}_6&=&\frac{1}{2048 (m_1^2 \sin ^2\alpha+m_2^2   \cos ^2\alpha)^3} 3 \cos^2\alpha ( \cos 2 \alpha  (8 \mu_2-3 \mu_1)+\mu_1-8 \mu_2)^2 \nonumber\\
&&\times (4 (m_1^2 \sin^2\alpha+m_2^2   \cos ^2\alpha) ( \cos 4 \alpha  (-6 \lambda_1-6 \lambda_2-2\lambda_3+6\lambda_4+3\lambda_5)\nonumber\\
&&+6 \lambda_1+6\lambda_2+2 \lambda_3+2\lambda_4+\lambda_5)- \cos ^2\alpha ( \cos2 \alpha  (3 \mu_1-8 \mu_2)-\mu_1+8 \mu_2) \nonumber\\
&&\times(3 \mu_1 \sin^2\alpha+8 \mu_2   \cos^2\alpha ))\nonumber\\
c^{GM}_8&=&\frac{1}{32768 (m
_1^2 \sin^2\alpha+m_2^2 \cos^2\alpha)^4} 3\cos^4\alpha ( \cos2 \alpha  (8\mu_2-3\mu_1)+\mu_1-8 \mu_2)^4 \nonumber\\
&&\times(3 (8 \lambda_1 \sin^4\alpha+\sin^2 2\alpha (2 \lambda_4+\lambda_5))+8 \cos^4\alpha (3\lambda_2+\lambda_3))\;.
\end{eqnarray}
In the small mixing limit, above relations reduce to
\begin{eqnarray}
c^{GM}_4&=&\lambda_1-\frac{3 \mu_1^2}{32 m_2^2}++\frac{\alpha ^2 (3 \mu _1^2 m_1^2+16 m_2^4 (-4 \lambda_1+2 \lambda_4+\lambda_5)+6 \mu_1 m_2^2 (3 \mu_1-8 \mu_2))}{32 m_2^4}+O(\alpha ^3) \nonumber\\
c^{GM}_6&=&\frac{3 \mu_1^2 (m_2^2 (2 \lambda_4+\lambda_5)-\mu_1 \mu_2)}{32 m_2^6}+\frac{3 \alpha ^2 \mu_1}{256 m_2^8} (24 \mu_1^2 \mu_2 m_1^2-\mu_1 m_2^2 (3 (\mu_1^2-24 \mu_1 \mu_2+64 \mu_2^2)\nonumber\\
&&+16 m_1^2 (2 \lambda_4+\lambda_5))+8 m_2^4 (\mu_1 (12 \lambda_1+12\lambda_2+4 \lambda _3-11 (2 \lambda_4+\lambda_5))+16 \mu_2 (2 \lambda_4+\lambda_5)))\nonumber\\
&&+O(\alpha ^3)\nonumber\\
c^{GM}_8&=& \frac{3 \mu_1^4 (3 \lambda_2+\lambda_3)}{256 m_2^8}+\frac{3 \alpha ^2 \mu_1^3}{512 \text{m2}^{10}}   (m_2^2 (3 \mu_1 (-24 \lambda_2-8 \lambda_3+2 \lambda_4+\lambda_5)+64 \mu_2 (3 \lambda_2+\lambda_3))\nonumber\\
&&-8 \mu_1 m_1^2 (3 \lambda_2+\lambda_3))+O(\alpha ^3)\;.
\end{eqnarray}

The cubic and quartic couplings relevant for the~\autoref{fig:h6dia} in xSM are 
\begin{eqnarray}
\lambda^{\rm xSM}_{2h2H}&&=\frac{1}{4} (3 (a_2-b_4-\lambda )\cos 4 \alpha  +a_2+3 (b_4+\lambda ))\;,\nonumber \\
\lambda^{\rm xSM}_{3H}&&=\frac{3}{2} a_1 \sin \alpha ^2 \cos \alpha +2 b_3 \cos \alpha^3 \;,\nonumber \\
\lambda^{\rm xSM}_{2hH}&&=\frac{1}{2} \cos \alpha (a_1   \cos \alpha ^2-2 (a_1-2b_3) \sin \alpha ^2) \;.
\end{eqnarray}

As in the GM model, one can use the equation of motion upon~\autoref{eq:xSMalp} to obtain the coefficients of $c_4$,$c_6$, $c_8$ in xSM, as follows
\begin{eqnarray}
c^{\rm xSM}_4&=&\frac{1}{4} (a_2 \sin^2\alpha \cos ^2\alpha+b_4 \sin ^4\alpha+\lambda \cos ^4\alpha)-\frac{  \cos ^2\alpha ((4 b_3-3 a_1) \cos 2 \alpha+a_1-4 b_3)^2}{128 (b_2 \cos^2\alpha-\mu ^2\sin^2\alpha)} \nonumber\\
c^{\rm xSM}_6&=&\frac{1}{1024 (b_2 \cos ^2\alpha-\mu ^2 \sin ^2\alpha)^3} \cos^2\alpha ((4b_3-3 a_1)  \cos 2 \alpha+a_1-4 b_3)^2 \nonumber\\
&&\times((3\cos4 \alpha (a_2-b_4-\lambda )+a_2+3 (b_4+\lambda )) (b_2 \cos ^2\alpha-\mu^2 \sin ^2\alpha)\nonumber\\
&&-\frac{1}{6} \cos\alpha ((3 a_1-4 b_3) \cos2 \alpha-a_1+4 b_3) (3 a_1  \sin ^2\alpha \cos \alpha+4 b_3   \cos ^3\alpha))\nonumber\\
c^{\rm xSM}_8&=&\frac{\cos ^4\alpha ((4 b_3-3 a_1) \cos2 \alpha+a_1-4 b_3)^4 (a_2 \sin ^2\alpha \cos^2\alpha+b_4 \cos ^4\alpha+\lambda \sin^4\alpha)}{16384 (b_2 \cos^2\alpha-\mu^2 \sin^2\alpha)^4}\;.
\end{eqnarray}
In the small mixing limit, one has
\begin{eqnarray}
c^{\rm xSM}_4&=&-\frac{a_1^2-8 b_2 \lambda }{32 b_2}+ \frac{\alpha ^2 (a_1^2 (6 b_2-\mu ^2)-8 a_1 b_2 b_3+8 b_2^2 (a_2-2 \lambda ))}{32 b_2^2} +O(\alpha ^3)\nonumber\\
c^{\rm xSM}_6&=& -\frac{a_1^2 (a_1 b_3-3 a_2 b_2)}{192 b_2^3}-\frac{\alpha^2 a_1}{256b_2^4}( a_1^3 b_2+4 a_1^2 b_3 (\mu^2-3b_2)\nonumber\\
&&+4 a_1 b_2 (a2 (11 b_2-2 \mu^2)-6 b_2 (b4+\lambda)+4 b_3^2)-32 a_2 b_2^2 b_3)+O(\alpha ^3)\nonumber\\
c^{\rm xSM}_8&=& \frac{a_1^4 b_4}{1024 b_2^4}+\frac{a_1^3 \alpha ^2}{1024 b_2^5} (a_1 (a_2 b_2+4 b_4 (\mu ^2-3b_2))+16 b_2 b_3 b_4)+O(\alpha ^3)
\end{eqnarray}
As studied in Ref.~\cite{Jiang:2018pbd}, the additional contribution to the $h^4$ can be reached, which captures the residual effects of the high dimension operators.  

\begin{figure}[!tbp]
\begin{center}
\includegraphics[width=\textwidth]{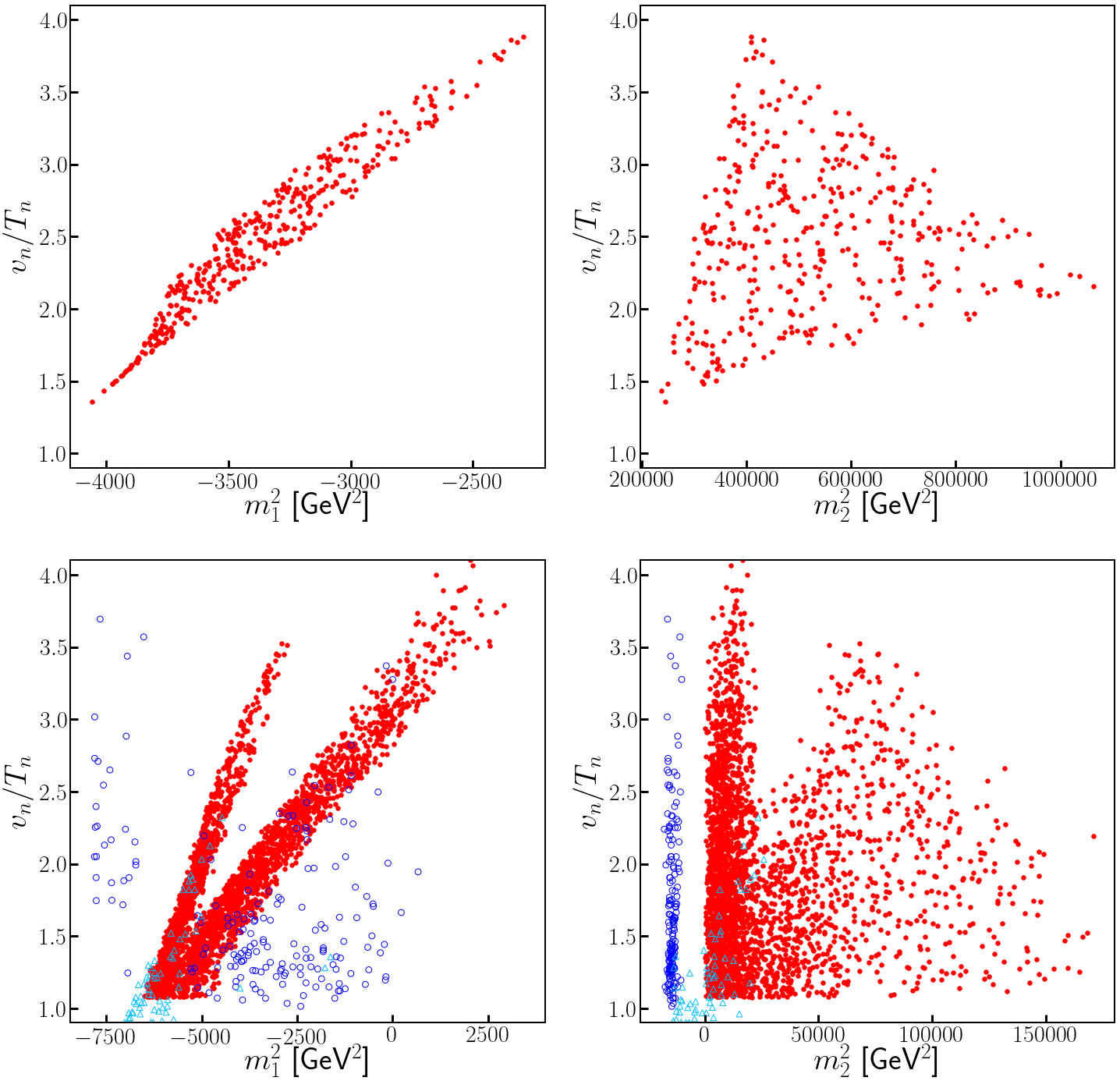}
\end{center}
\caption{ The $m_1^2$ (left panels) and $m_2^2$ (right panels) distributions for \hp{} (Top panels) and \lm{} (Bottom panels).}
\label{fig:12stepmass}
\end{figure}

\begin{figure}[!tbp]
\begin{center}
\includegraphics[width=\textwidth]{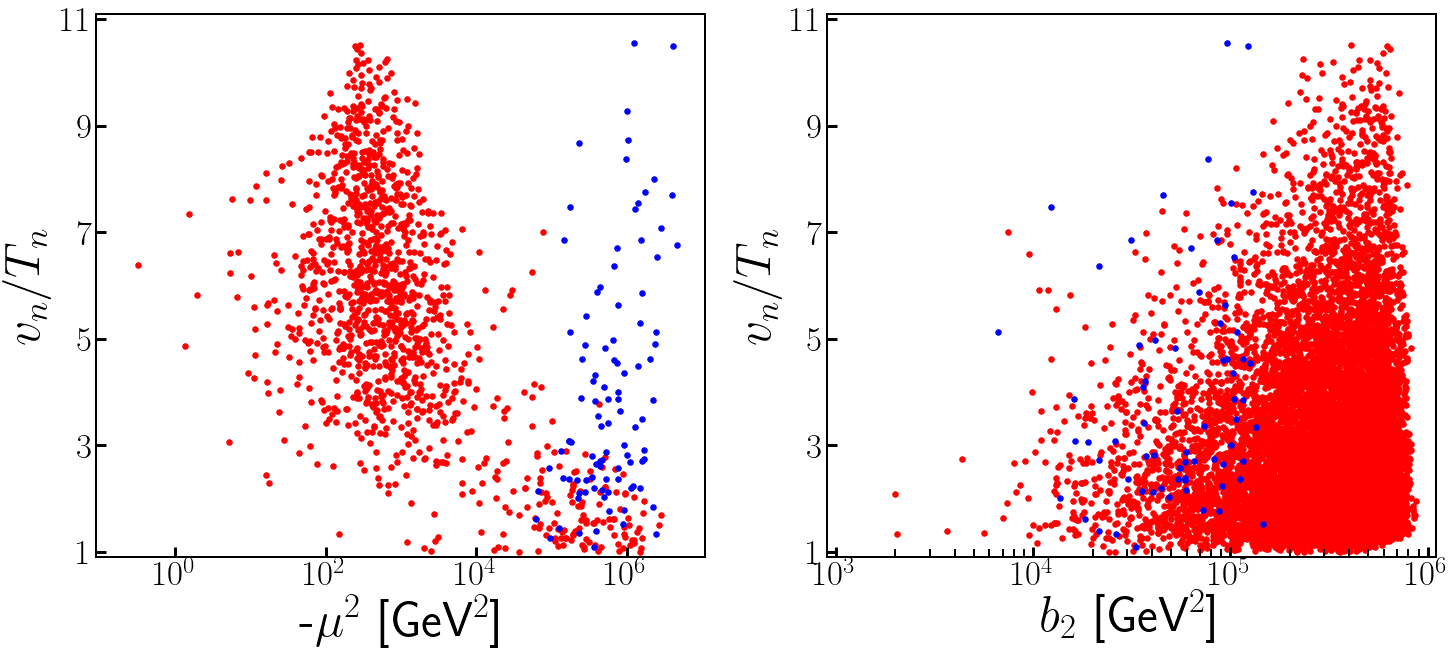}
\end{center}
\caption{ The $\mu^2$ and $b_2$ distributions of the one-step and two-step SFOEWPT valid points for xSM.}
\label{fig:12stepxsm}
\end{figure}

Up to now, we have calculated the high dimensional operators for the Higgs field by virtue of the method of CDE to match the UV models to the SMEFT~\cite{Henning:2014wua}. These high dimensional operators, especially the $(H^\dag H)^{3,4}$ are usually adopted to study phase transition in literatures.
However, according to~\autoref{fig:12stepmass} which shows the distributions of $m_1^2$ and $m_2^2$ in both \hp{} and \lm{}, the EFT approach does not apply for the GM model under study, especially the two-step case, as the basic assumption to integrate out the heavy degree of freedom is not fulfilled.
The corresponding cases in xSM are also demonstrated in~\autoref{fig:12stepxsm}.

The light degree freedom fields are usually necessary for the multi-step phase transition, see Ref.~\cite{Jiang:2015cwa,Cheng:2018axr,Bian:2018mkl,Bian:2018bxr,Cheng:2018ajh,Alves:2018oct}. In this situation, one could not employ the the effective field theory (EFT) approach explored in literatures (e.g., \cite{Grojean:2004xa,Huang:2015tdv}) to study the phase transition.
Even for the one-step phase transition, there are also some mismatch between the EFT approach and the UV complete model studies, especially for the scenario where not so small mixing among the SM Higgs and extra Higgs are present, see the study of Ref.~\cite{Damgaard:2015con} for the ``xSM" studies. We left the detailed survey on the match for the phase transition dynamics between UV complete models and EFT approach to further studies.  Indeed, the EFT approach for integrated out heavy degree freedoms valid for negligible mixing situations.

\section{Some Couplings in GM model}
\label{app:GMSVV}
\begin{align}
\begin{autobreak}
g_{H_{5}^{0} W^{-} W^{+} }=\frac{i e^2 v \sin \left(\theta _h\right) \bar{g}^{\rho \sigma }}{2 \sqrt{3} s_W^2}\;
	,
\end{autobreak}
\begin{autobreak}
g_{H_{5}^{0} Z Z }=-\frac{i e^2 v \sin \left(\theta _h\right) \bar{g}^{\rho \sigma }}{\sqrt{3} c_W^2 s_W^2}\;
	,
\end{autobreak}
\end{align}
\begin{align}
\begin{autobreak}
g_{H_{5}^{\mp} W^{\pm} Z }=-\frac{i e^2 v \sin \left(\theta _h\right) \bar{g}^{\rho \sigma }}{2 c_W s_W^2}\;
	,
\end{autobreak}
\begin{autobreak}
g_{H_{5}^{\mp\mp} W^{\pm} W^{\pm} }=\frac{i e^2 v \sin \left(\theta _h\right) \bar{g}^{\rho \sigma }}{\sqrt{2} s_W^2}\;
	,
\end{autobreak}
\end{align}
\begin{align}
\begin{autobreak}
g_{H W^{-} W^{+} }=\frac{i e^2 v \sin (\alpha ) \cos \left(\theta _h\right) \bar{g}^{\rho \sigma }}{2 s_W^2}
	+\frac{i \sqrt{\frac{2}{3}} e^2 v \cos (\alpha ) \sin \left(\theta _h\right) \bar{g}^{\rho \sigma }}{s_W^2}\;,
\end{autobreak}
\end{align}
\begin{align}
\begin{autobreak}
g_{H Z Z }=\frac{i e^2 v \sin (\alpha ) \cos \left(\theta _h\right) \bar{g}^{\rho \sigma }}{2 c_W^2 s_W^2}
	+\frac{i \sqrt{\frac{2}{3}} e^2 v \cos (\alpha ) \sin \left(\theta _h\right) \bar{g}^{\rho \sigma }}{c_W^2 s_W^2}\;,
\end{autobreak}
\end{align}
\begin{align}
\begin{autobreak}
g_{h W^{-} W^{+} }=\frac{i e^2 v \cos (\alpha ) \cos \left(\theta _h\right) \bar{g}^{\rho \sigma }}{2 s_W^2}
	-\frac{i \sqrt{\frac{2}{3}} e^2 v \sin (\alpha ) \sin \left(\theta _h\right) \bar{g}^{\rho \sigma }}{s_W^2}\;,
\end{autobreak}
\end{align}
\begin{align}
\begin{autobreak}
g_{h Z Z }=\frac{i e^2 v \cos (\alpha ) \cos \left(\theta _h\right) \bar{g}^{\rho \sigma }}{2 c_W^2 s_W^2}
	-\frac{i \sqrt{\frac{2}{3}} e^2 v \sin (\alpha ) \sin \left(\theta _h\right) \bar{g}^{\rho \sigma }}{c_W^2 s_W^2}\;,
\end{autobreak}
\end{align}

\begin{align}
\begin{autobreak}
g_{H_{3}^{0} h Z }=-\frac{\sqrt{\frac{2}{3}} e \sin (\alpha ) \cos \left(\theta _h\right)}{c_W s_W}
    -\frac{e \cos (\alpha ) \sin \left(\theta _h\right)}{2 c_W s_W}\;,
\end{autobreak}
\end{align}
\begin{align}
\begin{autobreak}
\lambda_{H h h }=
    -\frac{2 i \sin (\alpha ) \cos ^2(\alpha ) m_h^2 \sec \left(\theta _h\right)}{v}
    -\frac{2 i \sqrt{\frac{2}{3}} \sin (\alpha ) \sin (2 \alpha ) m_h^2 \csc \left(\theta _h\right)}{v}
    -\frac{i \sin (\alpha ) \cos ^2(\alpha ) m_H^2 \sec \left(\theta _h\right)}{v}
    -\frac{i \sqrt{\frac{2}{3}} \sin (\alpha ) \sin (2 \alpha ) m_H^2 \csc \left(\theta _h\right)}{v}
    -2 i \sqrt{2} \mu_1 \sin (\alpha ) \cos ^2(\alpha ) \cot \left(\theta _h\right)
    +i \sqrt{2} \mu_1 \sin ^3(\alpha ) \cot \left(\theta _h\right)
    -i \sqrt{3} \mu_1 \sin (\alpha ) \sin (2 \alpha ) \cot ^2\left(\theta _h\right)
    +i \sqrt{3} \mu_2 \sin (\alpha ) \sin (2 \alpha )\;,
\end{autobreak}
\end{align}

\begin{align}
\begin{autobreak}
\lambda_{h h h }=
    -\frac{3 i \cos ^3(\alpha ) m_h^2 \sec \left(\theta _h\right)}{v}
    +\frac{2 i \sqrt{6} \sin ^3(\alpha ) m_h^2 \csc \left(\theta _h\right)}{v}
    +3 i \sqrt{2} \mu_1 \sin ^2(\alpha ) \cos (\alpha ) \cot \left(\theta _h\right)
    +2 i \sqrt{3} \mu_1 \sin ^3(\alpha ) \cot ^2\left(\theta _h\right)
    -2 i \sqrt{3} \mu_2 \sin ^3(\alpha )\;.
\end{autobreak}
\end{align}

\newpage

\bibliographystyle{bibsty}
\bibliography{references}
\end{document}